\documentclass[times, 10pt,twocolumn]{article} 
\usepackage{latex8}
\usepackage{times}
\usepackage{graphicx}

\newcommand{\qed}{\hbox{\rlap{$~~\sqcap$}{$~~\sqcup$}}}
\newcommand{\itemS}{\vspace{-5pt} \item }

\newtheorem{algorithm}{Algorithm}
\newtheorem{theorem}{Theorem}

\newtheorem{definition}{Definition}
\newtheorem{corollary}{Corollary}

\begin{document}

\pagenumbering{roman}
\onecolumn
\begin{center}
\vspace{50pt}

{\Large Technical Report OBU-ECM-AFM-2019-01}
\vspace{50pt}

{\huge Morphy: A Datamorphic Software Test Automation Tool}

\vspace{20pt}

{\large Version 1.1, Dec. 14th, 2019}

\vspace{40pt}
{\large Hong Zhu, Ian Bayley}
\vspace{5pt}

School of Engineering, Computing and Mathematics\\
Oxford Brookes University\\
Oxford OX33 1HX, UK\\
Email: (hzhu$|$ibayley)@brookes.ac.uk
\vspace{20pt}

{\large Dongmei Liu, Xiaoyu Zheng}
\vspace{5pt}

School of Computer Science and Engineering

Nanjing University of Science and Technology

Nanjing 210094, China

Emails: dmliukz@njust.edu.cn, zxy961120@sina.com

\vspace{300pt}

{\large School of Engineering, Computing and Mathematics\\
Oxford Brookes University, Oxford OX33 1HX, UK}\\
\end{center}

\newpage

\begin{center}
\textbf{Abstract }
\end{center}

This paper presents an automated tool called $Morphy$ for datamorphic testing. It classifies software test artefacts into test entities and test morphisms, which are mappings on testing entities. In addition to datamorphisms, metamorphisms and seed test case makers, Morphy also employs a set of other test morphisms including test case metrics and filters, test set metrics and filters, test result analysers and test executers to realise test automation. In particular, basic testing activities can be automated by invoking test morphisms. Test strategies can be realised as complex combinations of test morphisms. Test processes can be automated by recording, editing and playing test scripts that invoke test morphisms and strategies. Three types of test strategies have been implemented in Morphy: datamorphism combination strategies, cluster border exploration strategies and strategies for test set optimisation via genetic algorithms. This paper focuses on the datamorphism combination strategies by giving their definitions and implementation algorithms. The paper also illustrates their uses for testing both traditional software and AI applications with three case studies. 
\newpage

\begin{center}
\tableofcontents
\contentsline {section}{{}References}{14}
\contentsline {section}{{}Appendix A. The Source code of TestCase and TestPool}{16}
\contentsline {section}{{}Appendix B. Example of Morphy Test Specification}{20}
\contentsline {section}{{}Appendix C. Template of Morphy Test Specification}{22}
\end{center}
\newpage

\twocolumn
\pagenumbering{arabic}

\title{Morphy: A Datamorphic Software Test Automation Tool}

\author{Hong Zhu, Ian Bayley\\
{\small School of Engineering, Computing and Mathematics}\\
{\small Oxford Brookes University}\\
{\small Oxford OX33 1HX, UK}\\
{\small Email: (hzhu$|$ibayley)@brookes.ac.uk}\\
\and
Dongmei Liu, Xiaoyu Zheng\\
{\small School of Computer Science and Engineering}\\
{\small Nanjing University of Science and Technology}\\
{\small Nanjing 210094, China}\\
{\small Emails: dmliukz@njust.edu.cn, zxy961120@sina.com}\\
}

\maketitle

\begin{abstract}
This paper presents an automated tool called $Morphy$ for datamorphic testing. It classifies software test artefacts into test entities and test morphisms, which are mappings on testing entities. In addition to datamorphisms, metamorphisms and seed test case makers, Morphy also employs a set of other test morphisms including test case metrics and filters, test set metrics and filters, test result analysers and test executers to realise test automation. In particular, basic testing activities can be automated by invoking test morphisms. Test strategies can be realised as complex combinations of test morphisms. Test processes can be automated by recording, editing and playing test scripts that invoke test morphisms and strategies. Three types of test strategies have been implemented in Morphy: datamorphism combination strategies, cluster border exploration strategies and strategies for test set optimisation via genetic algorithms. This paper focuses on the datamorphism combination strategies by giving their definitions and implementation algorithms. The paper also illustrates their uses for testing both traditional software and AI applications with three case studies. 
\end{abstract}


\Section{Introduction}

With the rapid growth of artificial intelligence (AI) in computer applications, ensuring the quality of software components that employ AI techniques becomes indispensable to software engineering. However, testing AI applications is notoriously difficult and prohibitively expensive \cite{AITest2019Proc}. It is highly desirable to advance software test automation techniques that meet the requirements of testing AI applications. 

Datamorphic testing has been proposed recently as an approach to software test automation \cite{Datamorphic2019}. In this method, test automation focuses on the development and application of three types of test code. 
\emph{Seed makers} generate test cases.
\emph{Datamorphisms} transform existing test cases into new ones. 
\emph{Metamorphisms} assert the correctness of test cases. 
Experiments \cite{Datamorphic2019,  DatamorphicTR2019, SalikThesis2019, YumiThesis2019} have demonstrated the it effective at testing AI applications. 

However, while datamorphic testing activities can be automated by writing project specific test code, it is highly desirable to develop a general testing tool to achieve the following requirements of test automation. 
\begin{enumerate}
\itemS \emph{Reusability} of test code of datamorphisms, metamorphisms, seed makers, etc, even across different projects. 
\itemS \emph{Composability } of test code in different combinations to conduct different experiments with the software under test. 
\itemS \emph{Constructability} of users' own test automation processes from existing test code so that the testing process can be repeated. 
\end{enumerate}

To achieve these goals, this paper extends the datamorphic testing framework by introducing the notion of test morphisms and presents an automated software test tool called \emph{Morphy}\footnote{Available at https://github.com/hongzhu6129/MorphyExamples.git}. It enables test automation at three levels. 
At the lowest level, various test activities can be performed by invoking test morphisms via a click of buttons on Morphy's graphic user interface. At the medium level, Morphy implements various test strategies to perform complicated testing activities through combinations and compositions of test morphisms. At the highest level, test processes are automated by recording, editing and replaying test scripts that consist of a sequence of invocations of test morphisms and strategies. 

The paper is organised as follows. Section \ref{sec:Sec2} extends the datamorphic testing framework. Section \ref{sec:MorphyTool}  presents the Morphy test tool. Section \ref{sec:Sec3} defines a set of strategies that combines datamorphisms to generate test data. Section \ref{sec:Sec4} reports three case studies to demonstrate the uses of Morphy. Section \ref{sec:Sec5} concludes the paper by a comparison with related work and a discussion of future work.

\Section{Extended Datamorphic Testing Framework} \label{sec:Sec2}

We extend datamorphic testing method by classifying the software artefacts involved in software testing into two kinds: \emph{entities} and \emph{morphisms}. 

\emph{Test entities} are objects and data used and/or generated in testing, which include \emph{test cases}, \emph{test suites}, the \emph{program under test}, and \emph{test reports}, etc. 

\emph{Test morphisms} are mappings between entities. They generate and transform test entities to achieve testing objectives. They can be implemented by writing test code. They can be invoked to perform test activities and composed to form test processes. Obviously, datamorphisms, metamorphisms and seed makers in the existing model of datamorphic testing are all test morphisms. However, there are other types of test morphisms that play crucial roles in test automation. 

A \emph{software test specification} in this extended framework specifies both of these artefacts and enables them to be invoked, composed as well as reused as a test library. In Morphy, a test specification is a Java class that declares a set of attributes for test entities and a set of methods for test morphisms. 

In this section, we discuss how they are defined in order to meet the requirements of test automation.

\SubSection{Test Entities}

Test cases and test suites are the most important kinds of entities on which test morphisms are defined. To enable the definition of various test morphisms, a test case must contain not only information about the input and output of the software, but also information about the following:
\begin{itemize}
\itemS How the test case is generated. Two particular pieces of information about the test case are recorded: whether it is a seed or a mutant, and which test morphism generates the test case. In the sequel, the former is called the \emph{feature} of the test case, the latter is called the \emph{type} of the test case. 
\itemS How a test case is \emph{related to} other test cases. If a test case is generated by using a datamorphism, the identities of test cases on which the datamorphism applied are recorded, and they are called the \emph{origins} of the test case. 
\itemS the \emph{correctness} of the test case. In datamorphic testing, the correctness of a test case is checked against metamorphisms. Each metamorphism can be a partial correctness condition. Therefore, test case may pass some of the metamorphisms but fail on the others. Therefore, the correctness of a test case is a set of records of checking the test case against metamorphisms. We will use the following format to record the correctness:  
 \[\{metamorphismName: (pass | fail)\}* \]
\end{itemize}

A test suite consists of a list of test cases. Each test case is also assigned with a universally unique identifier (UUID). Therefore, the relationships between test cases can be defined by references to their UUIDs. 

The Morphy testing tool defines two generic classes \emph{TestCase} and \emph{TestPool} for representing test cases and test suites, respectively. They have two type parameters for the input and output datatypes. 

The generic class \emph{TestCase} consists of attributes for (a) the UUID of the test case, (b) the input data, (c) the output data, (d) the feature, (e) the type of the test case, (f) the list of origins, and (g) the correctness of the test case. 

The generic class \emph{TestPool} consists of a list of TestCases and a number of methods for the operations of the test suite, such as adding and removing test cases to/from the test suite. The test suite used in the testing of the software is declared as an attribute of TestPool type and annotated with metadata \emph{@TestSetContainer}. A test specification class can also have attributes and methods without annotations. For examples, an attribute of TestPool type without annotation \emph{@TestSetContainer} can be used as an auxiliary test set.  

The source code of the TestCase and TestPool can be found in \emph{Appendix A}. 

\SubSection{Test Morphisms}

In addition to the three components of the original datamorphic testing model, we identify the following types of test morphisms that are useful to automate software testing.

\begin{itemize}
\itemS \emph{Test case metrics} are mappings from test cases to real numbers. They measure test cases, for example, on the similarity of a test case to the others in the test set. 
\itemS \emph{Test case filters} are mappings from test cases to truth values. They can be used, for example, to decide whether a test case should be included in the test set. 
\itemS \emph{Test set metrics} are mappings from test sets to real numbers. They measure the test set, for example, on its quality, such as its code coverage.  
\itemS \emph{Test set filters} are mappings from test sets to test sets. A typical example is to remove some test cases from a test set for regression testing. 
\itemS \emph{Test executers} execute the program under test on test cases and receive the outputs from the program. They are mappings from a piece of program to a mapping from input data to output. That is, they are functors in category theory. 
\itemS \emph{Test result analysers} analyse test results and generate test reports. Thus, they are mappings from test set to test reports. 
\end{itemize}

\SubSection{Test Specification}

A Morphy \emph{test specification} is a Java class, which declares a set of attributes as test entities and a set of methods as test morphisms; see \emph{Appendix B} for an example. Each test morphism is annotated with a metadata to declare the type of test morphism that the method belongs to. Table \ref{tab1} lists the annotations and datatypes of various types of test morphisms as implemented in Morphy. 

\vspace{-10pt}
\begin{table}[htbp]
\caption{Annotations of Test Morphisms}
{\footnotesize 
\begin{center}
\begin{tabular}{||l|l|l|l||}
\hline\hline
Morphism &Annotation &Parameter &Return\\
\hline\hline
Seed Maker &@SeedMaker &Nil &Void \\
\hline
Datamorphism &@Datamorphism &TestCase & TestCase\\
\hline
Metamorphism &@Metamorphism &TestCase &Boolean\\
\hline
Test Case Metrics &@TestCaseMetrics &TestCase &Real \\
\hline
Test Case Filter &@TestCaseFilter &TestCase &Boolean\\
\hline
Test Set Metrics &@TestSetMetrics &Nil &Real\\
\hline
Test Set Filter &@TestSetFilter &Nil & Nil\\
\hline
Test Executer &@TestExecuter &Input &Output \\
\hline
Analyser &@Analyser &Nil &Void\\
\hline
\hline
\end{tabular}
\end{center}
}
\label{tab1}
\end{table}
\vspace{-10pt}

The uses of various types of test entities and morphisms can be found in Section \ref{sec:Sec4}. 

\Section{Test Tool Morphy} \label{sec:MorphyTool}

\begin{figure}[htbp]
\centering
\scalebox{0.35}{\includegraphics{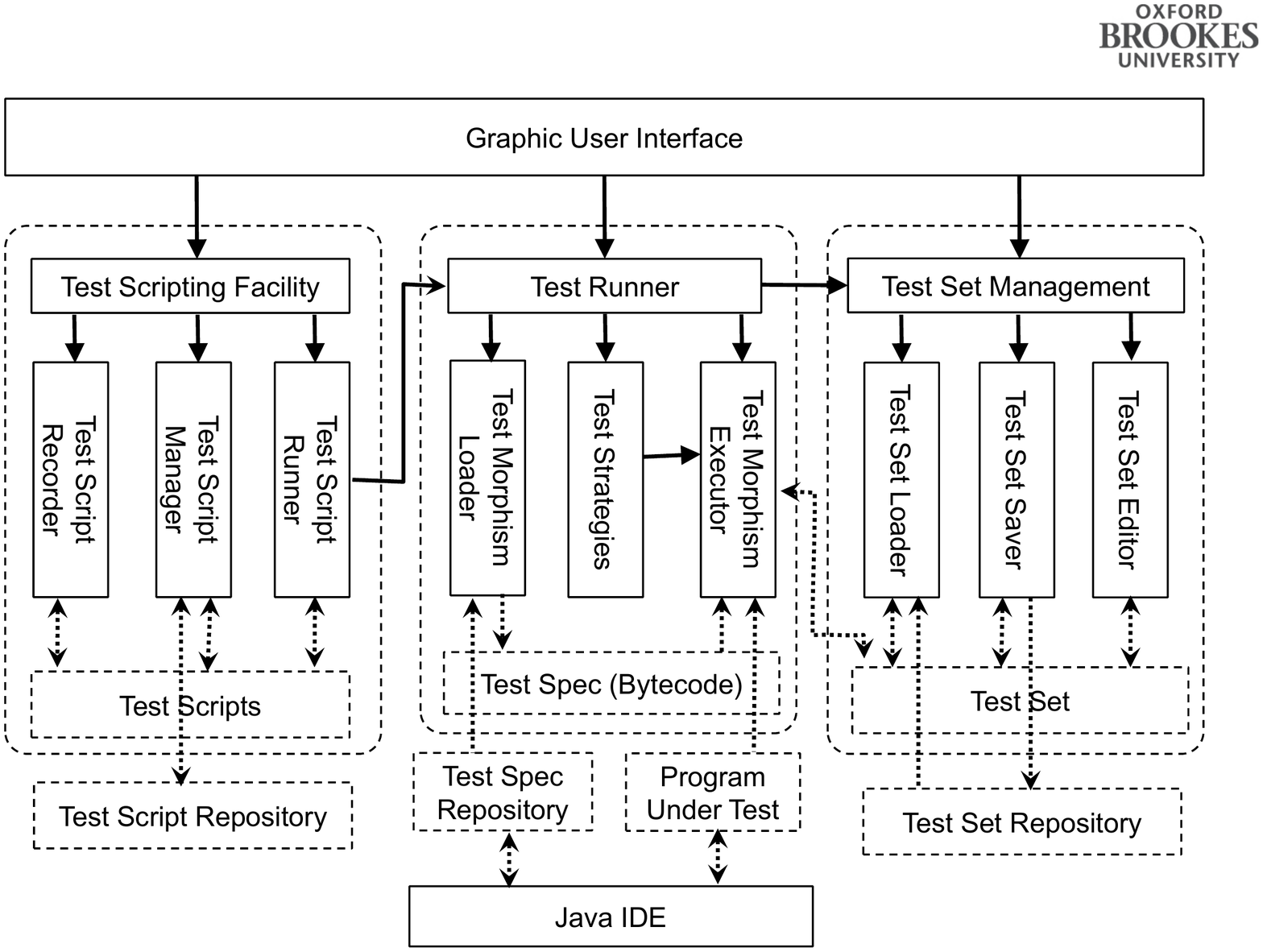}}
\caption{The Architecture of Morphy}
\label{fig:Fig0}
\end{figure}

As shown in Figure \ref{fig:Fig0}, Morphy consists of three main facilities: \emph{test set management}, \emph{test runner} and \emph{test scripting}. 

The test set management facility enables test sets to be saved into files, loaded from files and edited in a graphic user interface. 
The test runner enables test specifications to be loaded into the system and various test morphisms of the test specification to be invoked. It also implements various test strategies. The test scripting facility enables interactive testing activities to be recorded as a test script, saved into files, reloaded from files and replayed. 

Test specifications can be developed with any Java IDE, but a wizard has been developed as an Eclipse plugin to generate new skeleton Java class as a template of test specification. The skeleton code can be found in \emph{Appendix C}.  

\begin{figure*}[htbp]
\centering
\scalebox{0.21}{\includegraphics{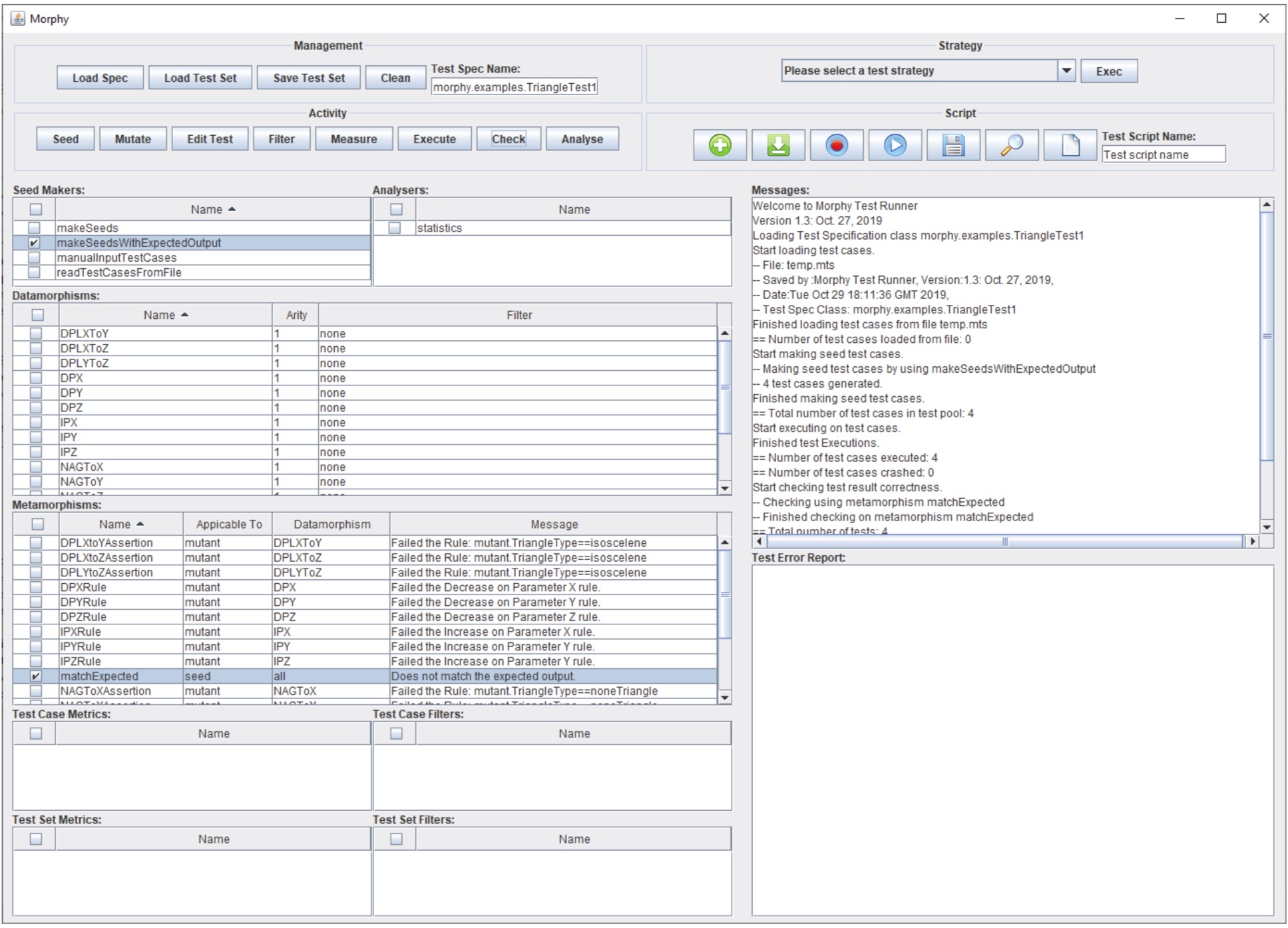}}
\caption{Morphy's Main GUI}
\label{fig:Fig1}
\end{figure*}

Morphy's main graphic user interface, shown in Figure \ref{fig:Fig1},  provides a user friendly environment in which testing artefacts can be managed, basic testing activities can be performed and automated testing facilities can be invoked. 

At the very top of Morphy's main window are four panels of buttons as follows. 

The \emph{Management} panel has buttons for the following functions to manage the artefacts of software testing; these include 
\begin{itemize} 
\itemS load a Morphy test specification, 
\itemS load a previously saved intermediate results of testing from a file, 
\itemS save the current state of the testing into a file,
\itemS clean up the system by removing all test cases, messages in the message areas, and the test scripts, etc. 
\end{itemize}
 
The \emph{Activity} panel has buttons for test activities that invoke various types of test morphisms. These include the following. 
\begin{itemize}
\itemS \emph{Seed}: generate seed test cases using selected seed maker methods; 
\itemS \emph{Mutate}: generate mutant test cases using selected datamorphisms; 
\itemS \emph{Filter}: remove test cases from the current test set using selected test set filters; 
\itemS \emph{Edit Test}: show the test cases in the current test pool in a test data window and to enable manual editing of the test results; 
\itemS \emph{Measure}: measure the current test set by invoking the selected test set metrics; 
\itemS \emph{Execute}: use the TestExecuter method to run testing on test cases in the current test set;
\itemS \emph{Check}: check the correctness of the test results using selected metamorphisms;
\itemS \emph{Analyse}: analyse the test results by invoking the selected analysis method(s).  
\end{itemize}

The \emph{Strategy} panel enables the tester to select from a set of predefined test strategies and to perform automated testing. Three types of test strategies have been implemented in Morphy: 
\begin{itemize} 
\itemS \emph{Mutant combination}: combining datamorphisms to generate mutant test cases; 
\itemS \emph{Domain exploration}: searching for the borders between clusters/subdomains of the input space; 
\itemS \emph{Test set optimisation}: optimising test sets by employing genetic algorithms. 
\end{itemize}
Due to the limitation of space, this paper focuses on the mutant combination strategies, which is given in Section \ref{sec:Sec3}. The other two types of test strategies will be reported separately. 

Finally, the buttons in the \emph{Test Script} panel enable the user to 
\begin{itemize} 
\itemS record testing activities as a test script, 
\itemS view the test script,
\itemS store a test script into a file, 
\itemS load a previously stored test script, and
\itemS play a test script. 
\end{itemize}

The test script facility allows the user to automate the testing process. It is particularly useful for repeated testing, such as in regression testing and to obtain data for statistical analysis. 

The left-hand side of the main window is a list of tables that shows various types of test morphisms.  The elements in these tables can be selected by clicking on the check boxes on the first column of the tables as input to perform the interactive and automatic testing functions mentioned above. 

The right-hand side contains two message panels. The upper one reports the testing activities performed and their outcomes. The lower report errors detected by checking the test results against metamorphisms. 

\begin{figure}[htbp]
\centering
\scalebox{0.21}{\includegraphics{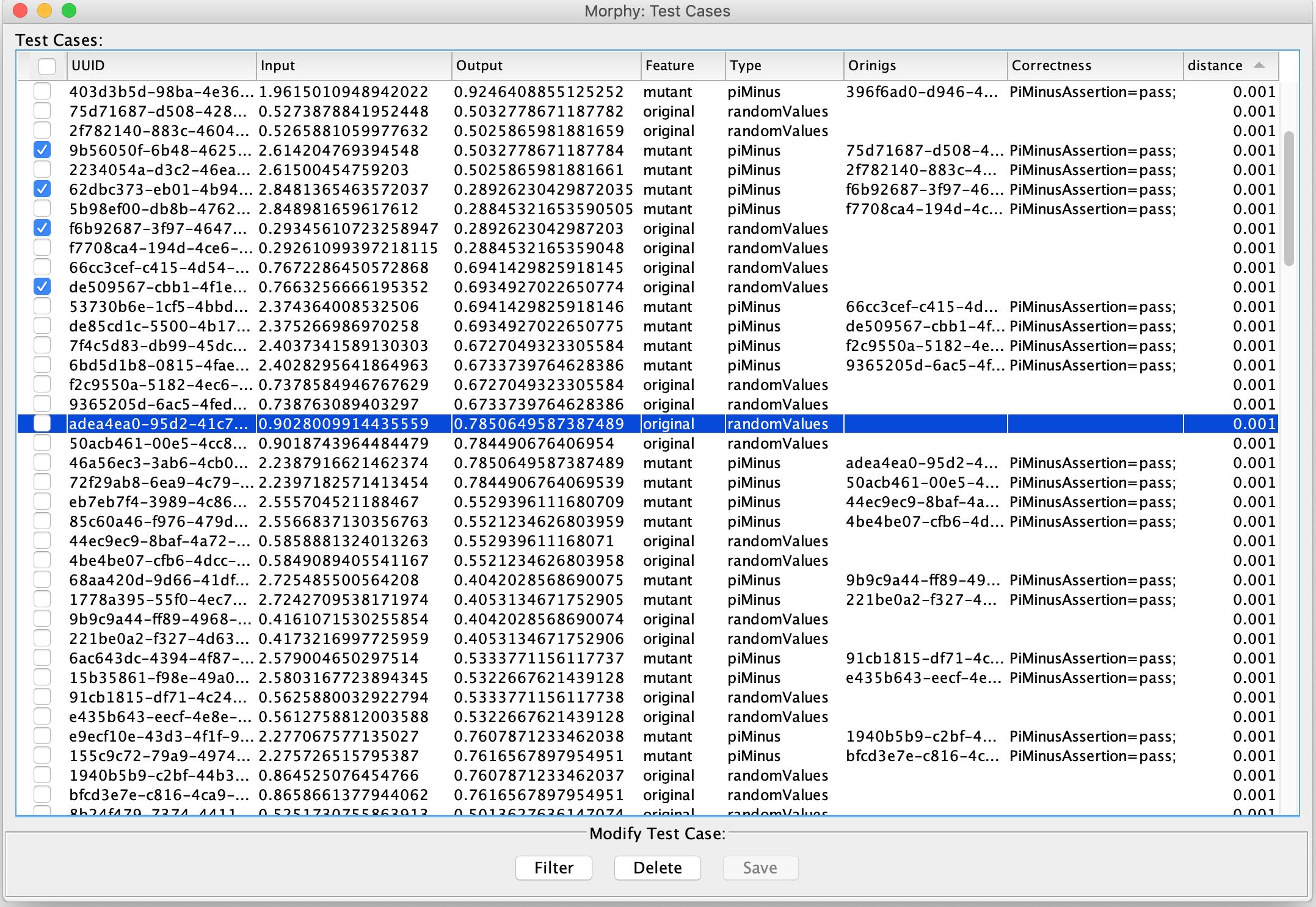}}
\caption{Morphy's Test Data Window}
\label{fig:Fig2}
\end{figure}

Figure \ref{fig:Fig2} shows Morphy's Test Case window. It shows the details of the test cases in the current test set. If test case metrics are defined in the test specification, the metrics will be applied to the test cases automatically and the results are shown in the table. The test cases in the table can be sorted by a click on the table head row. For example, the test cases in Figure \ref{fig:Fig2} are sorted according to the metrics Distance, shown in column Distance. Test cases can be selected by clicking on the check box on the first column. They can also be selected by applying a test case filter given in the test specification. When the tester clicks on the \emph{Delete} button, the selected test cases will be removed from the table. The \emph{Save} button will then be enabled. The deletion of the test cases from the test set will actually take place when the \emph{Save} button is clicked. 

\Section{Mutant Combination Strategies} \label{sec:Sec3}


Let $T$ be the set of all possible test cases for the software under test. $S \subset T$ be a set of test cases. $D$ be a set of datamorphisms and $d \in D$ is a datamorphism  in $D$. We say that $d$ is $k$-ary ($k>0$), if $d: T^k\rightarrow T$. 

\begin{definition} \label{def1} (First Order Mutants) 
A test case $y \in T$ is called a \emph{first order mutant test case}, or simply a \emph{first order mutant}, of $S$ generated by $D$, if there is a $k$-ary datamorphism $d \in D$ and test cases $x_1, \cdots, x_k \in S$  such that $ y = d(x_1, \cdots ,x_k)$. \qed
\end{definition}

\begin{definition} \label{def2} (First Order Mutant Completeness) 
A set $C$ of test cases is \emph{first order mutant complete} with respect to $S$ and $D$, if $S\subseteq C$, and  for each $d: T^k\rightarrow T \in D$, and each $x_i \in S$, $i=1,\cdots , k$, there is a test case $y \in C$ such that $y=d(x_1,x_2,\cdots ,x_k)$, where $d$ is $k$-ary. \qed
\end{definition}

In other words, a test set is first order mutant complete if it contains every seed and every first order mutant. A test strategy is to test the software with all the seeds and all the first order mutant test cases generated from the seeds using selected datamorphisms. 

The following algorithm generates the minimal test set that is first order mutant complete with respect to a give set of seed test cases and a set of datamorphisms.
 
\begin{algorithm} \label{alg1} (Generate 1st Order Mutant Complete Tests) 

{\footnotesize 
\begin{verbatim}
  Input: S = the set of seed test cases;
            D = the set of datamorphisms;
  Output: C = a set of test cases;
  Variables: tempT = temporal set of test cases;
  Begin
  1:   C = EmptySet; 
  2:   for (each datamorphism d in D){
  2.1:    tempT = EmptySet; 
  2.2:    Assume that d is a k-ary datamorphism;
  2.3:    forall k-tuples (x1,... ,xk) of S {
            add d(x1,... ,xk) to tempT;
          };
  2.4:    C = C + tempT; 
      };
  3:  return C + S;
  End
\end{verbatim}}\qed
\end{algorithm}

The following theorem proves the correctness of the algorithm. 

\begin{theorem}\label{thm1} The test set generated from $S$ using $D$ by Algorithm \ref{alg1} is the minimal set of test cases that is first order mutant complete with respect to $S$ and $D$. 
\end{theorem}
\vspace{-10pt}
\noindent{\emph{Proof.}} 

(a) \emph{Completeness}: Assume that the output test set $C$ from Algorithm 1 is not complete. This means there is either a seed test case $y$ not in $C$ or there is a first order mutant $y$ generated from seeds $x_1,\cdots ,x_k \in S$ by using a $k$-ary datamorphism $d \in D$ is not in $C$. In the former case, it is in conflict with Step 3. In the latter case, it is in conflict with Step 2.3. Therefore, the assumption is incorrect. 

(b) \emph{Minimalness}: It is obvious to see that the output only contains seeds and first order mutants. \qed

\begin{figure}[htbp]
\centering
\scalebox{0.18}{\includegraphics{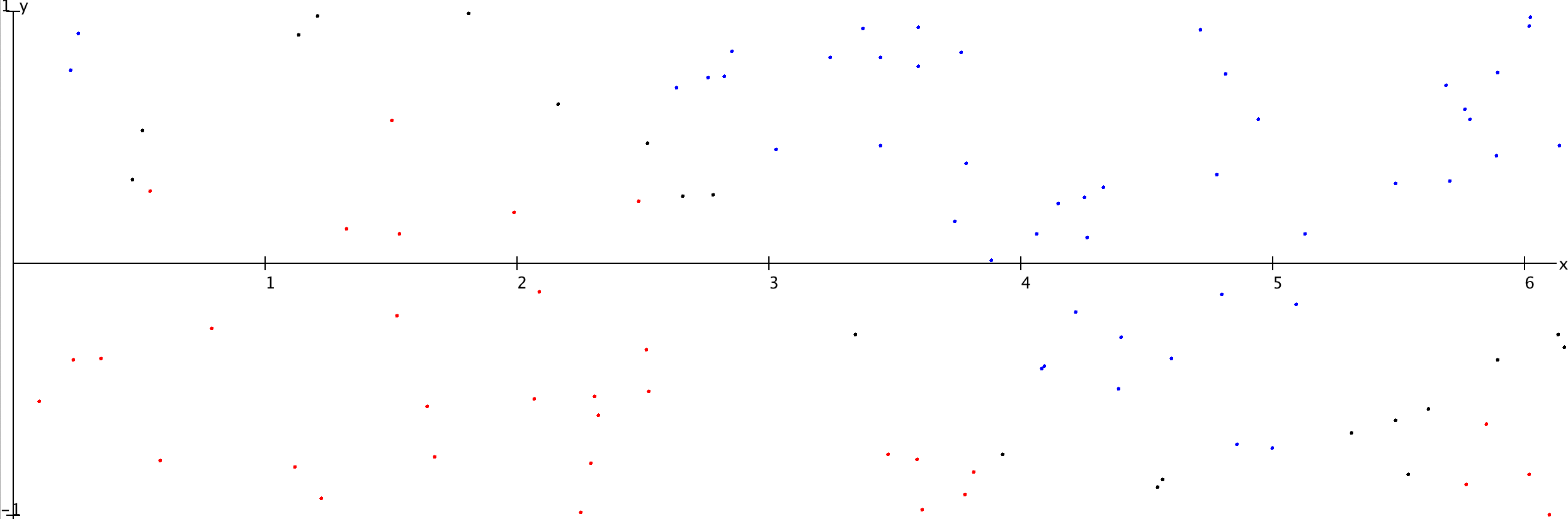}}

\footnotesize{(a) Original Test Set}

\scalebox{0.18}{\includegraphics{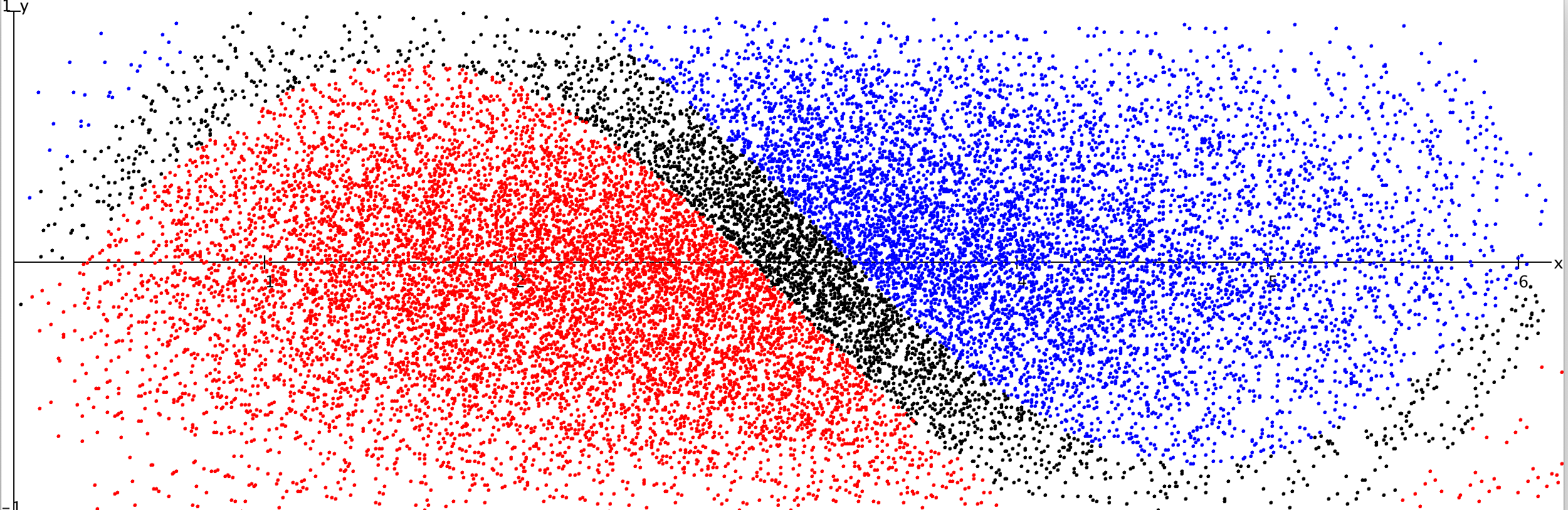}}
\footnotesize{(b) 1st Order Mutant Complete Test Set}
\caption{Test Results}
\label{fig:Fig3}
\end{figure}

For example, consider a software system that takes a point in the two-dimensional space of real numbers and classifies the points into three subdomains: the \emph{red}, the \emph{blue} and the \emph{black} areas. The test set initially contains 100 random points. The datamorphism is to add the middle point of two test cases. Applying Algorithm \ref{alg1} produces a first order mutant complete test set, which contains 10000 test cases. Figure \ref{fig:Fig3} (a) and (b) below shows the results of testing on the original 100 random test cases and on the 1st order mutant complete test set, respectively. 

Datamorphisms can also be applied to test cases multiple times to generate mutants of mutants, which are called \emph{high order mutants}. For the sake of convenience, a test case $x \in S$ is called a 0'th order mutant of $S$. 

\begin{definition} \label{def3} (Higher order mutants) 

A test case $y$ is a \emph{second order mutant} of $S$ by $D$, if there is a $k$-ary datamorphism $d \in D$ and $k$ test cases $x_1,\cdots, x_k$ such that $y = d(x_1,\cdots, x_k)$ and for all $x_i$, $x_i$ is either in $S$ or a first order mutant of $S$ by $D$, and at least one of $x_1,\cdots, x_k$ is a first order mutant of $S$ by $D$.  

A test case $y$ is an \emph{$n$'th order mutant} of $S$ by $D$ ($n>1$), if there is a $k$-ary datamorphism $d \in D$ and $k$ test cases $x_1,\cdots, x_k$ such that $y = d(x_1,\cdots, x_k)$ and $x_i$ are $m$'th order mutants of $S$ by $D$, where $m<n$, and at least one of $x_1,\cdots, x_k$ is a $(n-1)$'th order mutant of $S$ by $D$. \qed
\end{definition}

Similar to first order mutant completeness, a test set is 2nd order mutant complete if it contains all seed test cases, all 1st order mutants and all 2nd order mutant. In general, we have the following definition. 

\begin{definition} \label{def4} (K'th order mutant completeness) 
A set $C$ of test cases is $k$'th order mutant complete with respect to $S$ and $D$, if it contains all $i$'th order mutant test cases of $S$ by $D$ for all $i=0,\cdots, k$. \qed
\end{definition}

\begin{corollary} \textbf{\emph{of Theorem 1}}. 
By repeating Algorithm 1 for $K$ times that each time uses the output test set as the input to the next invocation of the algorithm, the result test set is the minimal $K$'th order mutant complete. 
\end{corollary}
\vspace{-10pt}
\noindent{\emph{Proof}. By induction on $K$. Details are omitted.} \qed

Assume that the set $D$ of datamorphisms contains $N$ methods. If a test set is $N$'th order mutant complete with respect to $S$ and $D$, it contains all permutations of the datamorphisms applied to all test cases. We say that the test set is \emph{permutation complete}. If the datamorphisms are associative, commutative, distributive and idempotent, a permutation complete test set contains all possible test cases that can be derived from a give set of test cases using the set of datamorphisms. The test set is therefore \emph{exhaustive} with regard to the set of seeds and the datamorphisms. It usually contains a huge number of test cases, so the costs of testing could be very high. A compromise is to cover the combinations of datamorphisms. 

A mutant of $S$ by $D$ can be represented as a tree on which the leaf nodes are test cases in $S$, and the non-leaf nodes are datamorphisms in $D$. The order of a mutant is the height of the tree. Figure \ref{fig:Fig4} below shows some examples of mutants, in which (a) and (b) are first order mutants, and (c) to (f) are second order mutants. 

\begin{figure}[htbp]
\centering
\scalebox{0.5}{\includegraphics{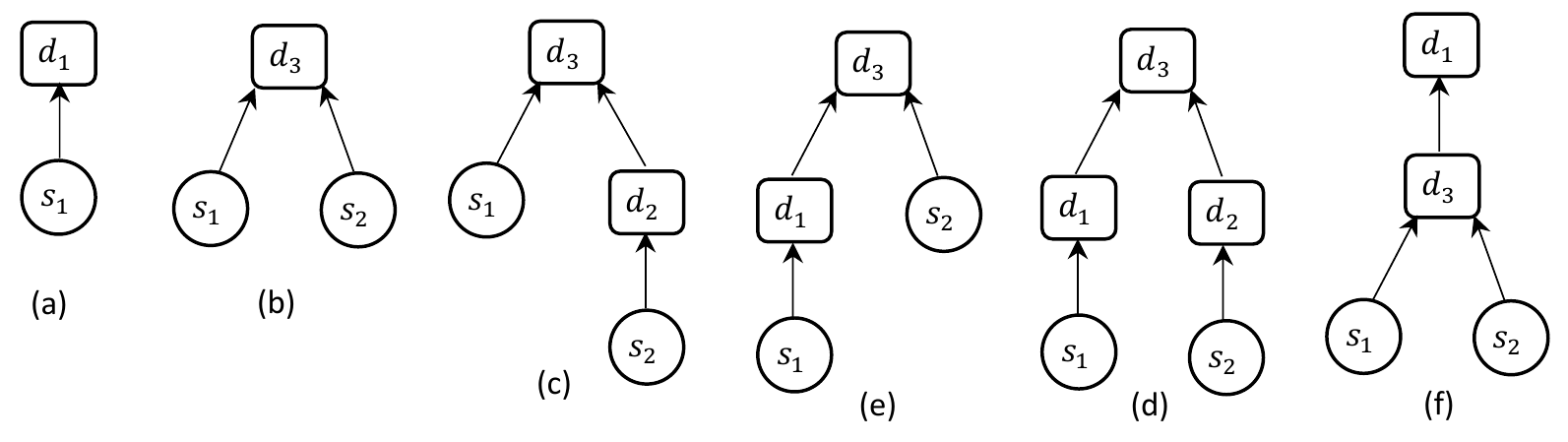}}
\caption{Examples of Mutant Trees}
\label{fig:Fig4}
\end{figure}

Given a mutant's tree representation, by replacing the test cases associated to the leaf nodes with variables in such a way that each different leaf node is associated with a different variable that ranges over the test cases, we can then obtain a function that generates a high order mutant when substitutes the variables with seed test cases. Each tree of this kind is therefore a way to combine datamorphisms to make higher order mutant test cases from seed test cases. We say that a combination $c$ is $k$-ary, if it contains $k$ variables, which is equivalent to the number of leaf nodes. We write $c(x_1,\cdots,x_k)$ to represent such a combination of datamorphism. When applying $c$ to seed test cases $a_1,\cdots,a_k$, we write $y = c(a_1,\cdots,a_k)$ to denote the result mutant test case. Let $\{d_1,\cdots,d_v \}$ be the set of datamorphisms in the tree, we also say that $c$ is a combination of $\{d_1,\cdots,d_v \}$. Given a set $D$ of datamorphisms, there may be many different combinations of $D$. 


\begin{definition} \label{def6} (Complete set of datamorphic combinations) A set $C$ of datamorphism combinations is \emph{combinatorial complete} for $D$, if for all subsets $D' \subseteq D$, there is a combination $c \in C$ that contains exactly the datamorphisms in $D'$.\qed
\end{definition}

\begin{definition} \label{def7} (Combinatorial complete test sets) A set $V$ of test cases is combinatorial complete with respect to $S$ and $D$, if
\begin{itemize}
\itemS there is a set $C$ of datamorphism combinations that is combinatorial complete with respect to $D$; and 
\itemS for every combination $c \in C$, if $c$ is $k$-ary, then for all $k$-tuple of test cases $(x_1,\cdots,x_k) \in S^k$, there is a test case $y$ in $V$ such that $y = c(x_1,\cdots,x_k)$. \qed
\end{itemize}
\end{definition}

The following is an algorithm that generates a combinatorial complete test set. 

\begin{algorithm} \label{alg2} (Generate Combinatorial Complete Test Set) 

{\footnotesize 
\begin{verbatim}
Input: S = the set of seed test cases;
          D = the set of datamorphisms;
Output: C = a set of test cases;
Variables: tempT = temporal set of test cases; 
Begin
1:     for (each datamorphism d in D) {
1.1:     tempT = empty_set; 
1.2:     Assume d is a k-ary, where k>0;
1.3:     for (all k-tuples (x1,...,xk) of S){
               add d(x1,...,xk) to tempT;
           };
1.4:     S = S + tempT; 
        };
2:     return C + S; 
End 
\end{verbatim}
}
\end{algorithm}

\begin{theorem}\label{thm2} 
The test set generated by Algorithm \ref{alg2} is combinatorial complete with respect to $S$ and $D$. 
\end{theorem}
\vspace{-10pt}
{\noindent \emph{Proof}.} 
Let $C$ be a subset of $D$. We write $||X||$ to denote the size of a set $X$. Let $||D||=m$. We proof by induction on $n=||C||\leq m$ that the test set generated by the algorithm is complete with respect to $C$. 

When $n=0$: the combination $c$ contains no datamorphism. By step 2, the result test set contains $S$. Thus, the statement is true in this case. 

Assume that when $n=N-1<m= ||D||$ the statement is true. 

Now let $n=N\leq m$, and $C$ be any given subset of datamorphisms and $||C||=N$. Let $d_1,d_2,\cdots,d_m$ be the sequence of datamorphisms executed in the order through the for-loop in Step 2. Let $\left<d_{i_1},\cdots,d_{i_N} \right>$ be the order that datamorphisms in $C$ occurred in the for-loop. By the induction assumption, there is a datamorphism combination $c'$ that contains all datamorphisms $d_{i_1}, \cdots, d_{i_{(N-1)}}$. Therefore, by Step 2.3 of the algorithm, all permutations of the test cases of the $c'$ as well as all other test cases, including seeds, will be used to generate test cases $d_{i_N} (x_1,\cdots,x_k)$. Therefore, the statement is also true for $n=N \leq m$. 

By the induction principle, the theorem is true. \qed

Note that, the test set generated by Algorithm \ref{alg2} may be not minimal in size if there is a datamorphism that is non-unary.  

\section{Case Studies} \label{sec:Sec4}

In this section, we report three case studies on the uses of Morphy in automated software testing. \footnote{The source code of the case studies can be found on GitHub at the URL: https://github.com/hongzhu6129/MorphyExamples.git}

\subsection{Triangle Classification}

Triangle classification is a classic software testing problem that Myer used to illustrate the importance of combination of various types of test cases \cite{MyerBook2004}. The program under test \emph{``reads three integer values from an input dialog. The three values represent the lengths of the sides of a triangle. The program displays a message that states whether the triangle is scalene, isosceles, or equilateral."} \cite{MyerBook2004}
Myer listed 14 questions for testers to assess how well he/she tests the program for such a seemly simple program and reported that \emph{``highly qualified professional programmers score, on the average, only 7.8 out of a possible 14"}. Here, it is used to demonstrate how to design datamorphisms according to Myer's assessment criteria to achieve test adequacy and how datamorphic testing method automates the testing process.

\subsubsection{The input and output datatypes}

Slightly different from the original problem that Mayer specified, the program under test is implemented as a Java class with three attributes $x$, $y$ and $z$ of integer type and a method \emph{classify} to classify the triangle into four types: $equilateral$, $isosceles$, $scalene$, $noneTriangle$. Therefore, the input data type is a class called \emph{Triangle} with two methods: \emph{toString()} and \emph{valueOf()}, to convert between the object and string representations of a test case. A Java enum class \emph{TriangleType} is declared as the types of triangles. It is the output datatype of test cases. 

%
%

\subsubsection{Generation of seed test cases}

There are a number of ways to generate seed test cases in the datamorphic approach. This case study shows how each of them can be easily implemented in a Morphy test specification. 

\noindent{ \emph{A. Literal Constants}. The first is to generate seed test cases as hard coded literal constants and save them into the test pool. The following method in the TriangleTestSpec class is a seed maker morphism that generates 4 seed test cases, one in each typical type of triangles. 
}

\begin{footnotesize} 
\begin{verbatim}
  @MakeSeed
  public void makeSeeds(){
    testSuite.addInput(new Triangle(5,5,5));
    testSuite.addInput(new Triangle(5,5,7));
    testSuite.addInput(new Triangle(5,7,9));
    testSuite.addInput(new Triangle(3,5,9));
  }
\end{verbatim}
\end{footnotesize} 

{\noindent \emph{B. Constants with Expected Output}. Expected outputs on seed test cases can also be stored and results of test executions can be checked against the expected outputs. To do so, an additional test pool can be declared to store the expected answers to the seed test cases. The following is a segment of the code in Java. 
}

\begin{footnotesize} 
\begin{verbatim}
  public TestPool<Triangle,TriangleType> expected 
     = new TestPool<Triangle, TriangleType>();
  
  @MakeSeed
  public void makeSeedsWithExpectedOutput(){
    Triangle trg;
    TestCase<Triangle, TriangleType> tc;
    
    trg = new Triangle(5,5,5); 
    tc = new TestCase<Triangle, TriangleType>(); 
    tc.input = trg;
    tc.setFeature(TestDataFeature.original);
    testSuite.addTestCase(tc);
    tc.output = TriangleType.equilateral;
    expected.addTestCase(tc);
  }
\end{verbatim}
\end{footnotesize} 

A metamorphism can be defined to check the correctness of test executions on seed test cases. The following is such an example. 

\begin{footnotesize} 
\begin{verbatim}
  @Metamorphism(
    applicableTestCase="seed",
    message="Match the expected output."
  )
  public boolean matchExpected(
       TestCase<Triangle, TriangleType> x) {
    String tcId = x.id;
    TestCase expectedTc = expected.get(tcId);
    return (x.output == expectedTc.output); 
  }
\end{verbatim}
\end{footnotesize} 

{\noindent \emph{C. Manual Input}. Seed test cases can be entered into the system through interactive manual input. The following is such an example code. 
}
\begin{footnotesize} 
\begin{verbatim}
  @MakeSeed
  public void manualInputTestCases(){
    Triangle trg;
    TestCase<Triangle, TriangleType> tc; 
    String numStr; 
    while (true) {
      tc = new TestCase<Triangle,TriangleType>();
      trg = new Triangle(); 
      numStr = JOptionPane.showInputDialog(
        "Please input x:");
      if (numStr==null) { break;}
      trg.x  = Integer.valueOf(numStr);
      numStr = JOptionPane.showInputDialog(
        "Please input y:");
      if (numStr==null) { break;}
      trg.y  = Integer.valueOf(numStr);
      numStr = JOptionPane.showInputDialog(
        "Please input z:");
      if (numStr==null) { break;}
      trg.z  = Integer.valueOf(numStr);
      tc.input = trg;
      tc.setFeature(TestDataFeature.original);
      testSuite.addTestCase(tc);
    }
  }
\end{verbatim}
\end{footnotesize} 

{\noindent \emph{D. Read Test Cases from A File}. Test cases can also read from a file and then added to the test suite. 
For the sake of space, here the code is omitted.
}
%

\subsubsection{Datamorphisms}

The main body of the test specification consists of a set of datamorphisms and metamorphisms. The datamorphisms implement Mayer�s test requirements. For example, Mayer requires that a test set contains all permutations of the three input values so a set of datamorphisms have been designed to generate mutants that are the permutations of the seed test cases. The following is one such datamorphism.

\begin{footnotesize} 
\begin{verbatim}
  @Datamorphism
  public TestCase<Triangle, TriangleType> swapXY(
      TestCase<Triangle, TriangleType> seed){
    TestCase<Triangle, TriangleType> mutant 
      = new TestCase<Triangle, TriangleType>();
    Triangle m = new Triangle(1,1,1);
    m.x=seed.input.y;
    m.y=seed.input.x;
    m.z=seed.input.z;
    mutant.input = m;
    return mutant;
  }
\end{verbatim}
\end{footnotesize} 

Table 1 lists the datamorphisms contained in the test specification. As shown in the above example, these datamorphisms are very simple Java code; each is no more than 10 lines. 

\begin{table}[ htbp]
{\footnotesize 
\caption{List of Datamorphisms}
\begin{center}
\begin{tabular}{||l|l||}
\hline\hline
Name&Function\\
\hline\hline
increaseX &Increase the value of x by 1\\
\hline
increaseY &Increase the value of y by 1\\
\hline
increaseZ &Increase the value of z by 1\\
\hline
decreaseX &Decrease the value of x by 1\\
\hline
decreaseY &Decrease the value of y by 1\\
\hline
decreaseZ &Decrease the value of z by 1\\
\hline
swapXY &Swap the values of x and y\\
\hline
swapXZ &Swap the values of x and z\\
\hline
swapYZ &Swap the values of y and z\\
\hline
rotateL &Rotate the values of x, y and z left\\
\hline
rotateR &Rotate the values of x, y and z right\\
\hline
copyXToY &Copy the value of x to y\\
\hline
copyXToZ &Copy the value of x to y\\
\hline
copyYToZ &Copy the value of y to z\\
\hline
negateX &Negate the value of x\\
\hline
negateY &Negate the value of y\\
\hline
negateZ &Negate the value of z\\
\hline
zeroX &Set the value of x to 0\\
\hline
zeroY &Set of value of y to 0\\
\hline
zeroZ &Set of value of z to 0\\
\hline\hline
\end{tabular}
\end{center}
}
\label{default}
\end{table}%

Applying these datamorphisms with the first order mutant complete strategy generated 80 mutant test cases, which together with the seed test cases achieved the full coverage of Myer's test requirements. 

\subsubsection{Metamorphisms}

For each datamorphism, there is a corresponding metamorphism that makes an assertion about the expected output of the program on mutant test cases. For example, for datamorphism $swapXY$, which swaps the values of $x$ and $y$ of a triangle, the following metamorphism asserts that such a swap will not change the classification outcome. 

\begin{footnotesize} 
\begin{verbatim}
  @Metamorphism(
      applicableTestCase="mutant",
      applicableDatamorphism = "swapXY",
      message="Failed the Swap X Y rule."
      )
  public boolean swapXYRule(
      TestCase<Triangle, TriangleType> x) {
    String originalId = x.getOrigins().get(0);
    TestCase origTc=testSuite.get(originalId);
    return (origTc.output == x.output); 
  }
\end{verbatim}
\end{footnotesize} 

It is worth noting that annotatations restrict the mutant type to which a metamorphism can be applied. For example, the above metamorphism only applies to mutant test cases that are generated by applying the $swapXY$ datamorphism. 

The metamorphisms in this case study are also very simply; each has no more than 13 lines. 

\subsubsection{Test Execution and Result Analysis}

The test executions of the program under test can easily be defined by a test executer morphism. 
\begin{footnotesize} 
\begin{verbatim}
@TestExecuter
public TriangleType Classifier(Triangle tc) {
  Triangle1 x = new Triangle1(tc.x, tc.y, tc.z); 
  return x.Classify();
}
\end{verbatim}
\end{footnotesize} 

 
The analysis of test results can be performed by invoking an analyser morphism. In the case study, an analyser method is written for statistical analysis of test cases and it reports the data in a pop-up. The details are omitted for the sake of space.

\begin{figure}[htbp]
\centering
\scalebox{0.545}{\includegraphics{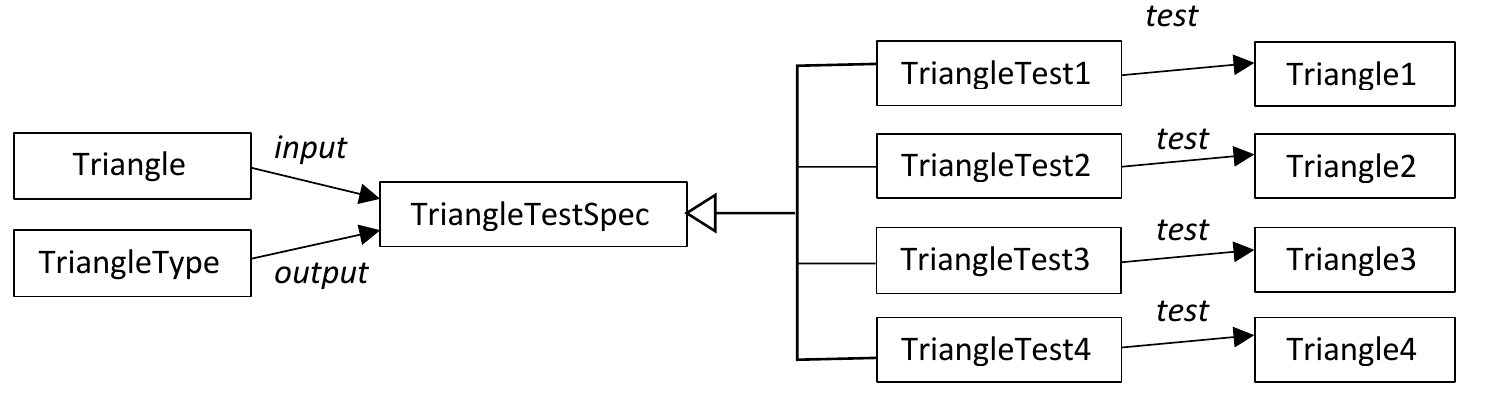}}
\caption{Classes of Test Specification}
\label{fig:Fig5}
\end{figure}

Note that test entities and morphisms can be declared in a number of classes to make them more reusable. For example, in this case study, we have put the test executer method in a separate class that inherits the class $TriangleTestSpec$, where the datamorphisms and metamorphisms are declared. Consequently, the test specification class $TriangleTestSpec$ can be reused to test a number of different implementations and thier versions even if their interface is different. In the case study, we tested two different algorithms for triangle classification, and for each of them, we made two versions, one with errors and one without. The relationships between various classes used in the case study is depicted in Figure \ref{fig:Fig5}. 

Using such an organisation for test specifications not only reuses the classes, but also make test scripts reusable without change for testing various different implementations. 

\subsection{Trigonometric Functions}

In this case study, we test three trigonometric functions $sin(x)$, $cos(x)$ and $tan(x)$ provided by Java Math library. The correctness of the library�s implementation of these functions will be checked against a set of trigonometric identities and on a set of special values between 0 and $2\pi$ as well as random test cases. 

\subsubsection{Test Cases and Test Suite}

The inputs to these trigonometric functions are real numbers and so are the results. However, instead of using $double$ or $Double$ as the output datatype of test cases, we declare a class called $Trigonometrics$ that contains three attributes $sin$, $cos$ and $tan$ to store the values of these functions and use it as the output datatype. This enables us to check the functions on identities that involves multiple trigonometric functions. 
%
%
%
%
%

Two seed maker morphisms are included in the test specification. One generates 20 random inputs in the range from 0 to $\frac{\pi}{2}$, and the other generates a set of special values and the expected output of the functions; see      
Table \ref{tab:Tab3}. 

\begin{table*}[ htbp]
\caption{The Special Values of Trigonometric Functions Used as Special Test Cases} \label{tab:Tab3}
\begin{center}
{\footnotesize
\begin{tabular}{|c|c|c|c|c|c|c|c|c|c|c|c|c|c|c|c|c|c|}
\hline 
$x$ &$0$ &$\frac{\pi}{6}$ &$\frac{\pi}{4}$ &$\frac{\pi}{3}$ &$\frac{\pi}{2}$ &$\frac{2\pi}{3}$ &$\frac{3\pi}{4}$ &$\frac{5\pi}{6}$ &$\pi$ &$\frac{7\pi}{6}$ &$\frac{5\pi}{4}$ &$\frac{4\pi}{3}$ &$\frac{3\pi}{2}$ &$\frac{5\pi}{3}$ &$\frac{7\pi}{4}$ &$\frac{11\pi}{6}$ &$2\pi$\\
\hline
$sin(x)$ &$0$ &$\frac{1}{2}$ &$\frac{\sqrt{2}}{2}$ &$\frac{\sqrt{3}}{2}$ &$1$ &$\frac{\sqrt{3}}{2}$ &$\frac{\sqrt{2}}{2}$ &$\frac{1}{2}$ &$0$ &$-\frac{1}{2}$ &$-\frac{\sqrt{2}}{2}$ &$-\frac{\sqrt{3}}{2}$ &$-1$ &$-\frac{\sqrt{3}}{2}$ &$-\frac{\sqrt{2}}{2}$ &$-\frac{1}{2}$ &$0$\\
\hline
$cos(x)$ &$1$ &$\frac{\sqrt{3}}{2}$ &$\frac{\sqrt{2}}{2}$ &$\frac{1}{2}$ &$0$ &$-\frac{1}{2}$ &$-\frac{\sqrt{2}}{2}$ &$-\frac{\sqrt{3}}{2}$ &$-1$ &$-\frac{\sqrt{3}}{2}$ &$-\frac{\sqrt{2}}{2}$ &$-\frac{1}{2}$ &$0$ &$\frac{1}{2}$ &$\frac{\sqrt{2}}{2}$ &$\frac{\sqrt{3}}{2}$ &$1$\\
\hline
$tan(x)$ &$0$ &$\frac{\sqrt{3}}{3}$ &$1$ &$1$ &$\infty$ &$-\sqrt{3}$ &$-1$  &$-\frac{\sqrt{3}}{3}$ &$0$ &$\frac{\sqrt{3}}{3}$ &$1$ &$\sqrt{3}$ &�$\infty$ &$-\sqrt{3}$ &$-1$  &$-\frac{\sqrt{3}}{3}$ &$0$\\
\hline
\end{tabular}
}
\end{center}
\end{table*}%

\subsubsection{Datamorphisms and Metamorphisms}

The datamorphisms for testing trigonometric functions are very simple functions on real numbers; see Table \ref{tab:Tab4}. A set of identities of trigonometric functions are implemented as metamorphisms; see Table \ref{tab:Tab5}. 

\begin{table}[h]
\caption{List of Datamorphisms} \label{tab:Tab4}
\begin{center}
{\footnotesize
\begin{tabular}{||l|l||l|l||}
\hline\hline
Name &Function &Name & Function\\
\hline\hline
 halfPiPlus &$x \rightarrow \pi/2 + x$ &halfPiMinus &$x \rightarrow \pi/2 ? x$ \\
 \hline
 piPlus &$x \rightarrow \pi + x$  &piMinus &$x \rightarrow \pi ? x$  \\
 \hline
 twoPiPlus &$x \rightarrow 2\pi + x$ &twoPiMinus &$x \rightarrow  2\pi ? x$\\
 \hline
 sum & $(x,y) \rightarrow  x + y$  &diff & $(x,y) \rightarrow x ? y$  \\
 \hline
 negate &$x \rightarrow-x$ &&\\
 \hline\hline
\end{tabular}
}
\end{center}
\end{table}%

\begin{table}[h]
\caption{List of Metamorphisms} \label{tab:Tab5}
\begin{center}
{\footnotesize
\begin{tabular}{ll}
\hline
$sin(\pi-x) = sin(x)$  &$sin(\pi + x) = -sin(x)$  \\
$cos(\pi-x) = -cos(x)$  &$cos(\pi + x) = -cos(x)$ \\
$tan(\pi-x) = -tan(x)$ &$tan(\pi+x) = tan(x)$ \\
$sin(\pi/2+x) = cos(x)$ &$sin(\pi/2-x) = cos(x)$ \\
$cos(\pi/2+x) = -sin(x)$  &$cos(\pi/2-x) = sin(x)$ \\
$tan(\pi/2+x) = -1/tan(x)$ &$tan(\pi/2-x) = 1/tan(x)$\\
$sin(2\pi-x) = -sin(x)$ &$sin(2\pi+x) = sin(x)$ \\
$cos(2\pi-x) = cos(x)$ &$cos(2\pi+x) = cos(x)$ \\
$tan(2\pi-x) = -tan(x)$ &$tan(2\pi+x) = tan(x)$\\
$sin(-x) = -sin(x)$ &$cos(-x) = cos(x)$\\ 
\end{tabular}
\begin{tabular}{l}
$tan(-x) = -tan(x)$\\
$sin(x+y) = sin(x)cos(y)+cos(x)sin(y)$ \\
$cos(x+y) = cos(x)cos(y) - sin(x)sin(y)$ \\
$sin(x-y) = sin(x)cos(y)-cos(x)sin(y)$ \\
$cos(x-y) = cos(x)cos(y) + sin(x)sin(y)$\\
$tan(x+y) =  (tan(x)+tan(y))/(1-tan(x)tan(y))$ \\
$tan(x-y) =  (tan(x)-tan(y))/(1+tan(x)tan(y))$\\
\hline
\end{tabular}
}
\end{center}
\end{table}%

The implementations of metamorphisms are straightforward. The following is an example. 
\begin{footnotesize} 
\begin{verbatim}
  double error = 0.000000000001;
  @Metamorphism(
      applicableTestCase="mutant",
      applicableDatamorphism="HalfPiMinus",
      message="The rule: sin(pi/2-x) = cos(x)"
  )
  public boolean HalfPiMinusSinAssertion(
       TestCase<Double, Trigonometrics> tc) {
    TestCase<Double, Trigonometrics> originalTc 
       = testSuite.get(tc.getOrigins().get(0));
    return (Math.abs(tc.output.sin 
      - originalTc.output.cos) <= error);
  }
\end{verbatim}
\end{footnotesize} 

\subsubsection{Test Executions and Analysis of Results}

The execution of the program on test cases is defined by the following test executer method, which invokes the program under test and stores the output of the program to the test case. 

\begin{footnotesize} 
\begin{verbatim}
  @TestExecuter
  public Trigonometrics testMath(Double x) {
    Trigonometrics result = new Trigonometrics();
    result.sin = Math.sin(x);
    result.cos = Math.cos(x);
    result.tan = Math.tan(x);
    return result;
  }
\end{verbatim}
\end{footnotesize} 

Two analyser methods were used: one for statistical analysis of the test results, reusing the analyser for testing triangle classification program, and the other for visualising the test outputs. 

The testing process consists of two stages. The first stage starts with the generation of special value test cases, executions of the functions on them and then using a metamorphism to check whether the output matches the expected value. The analyse of the test results shows that the test detected no error.

\begin{figure}[htbp]
\centering
\scalebox{0.12}{\includegraphics{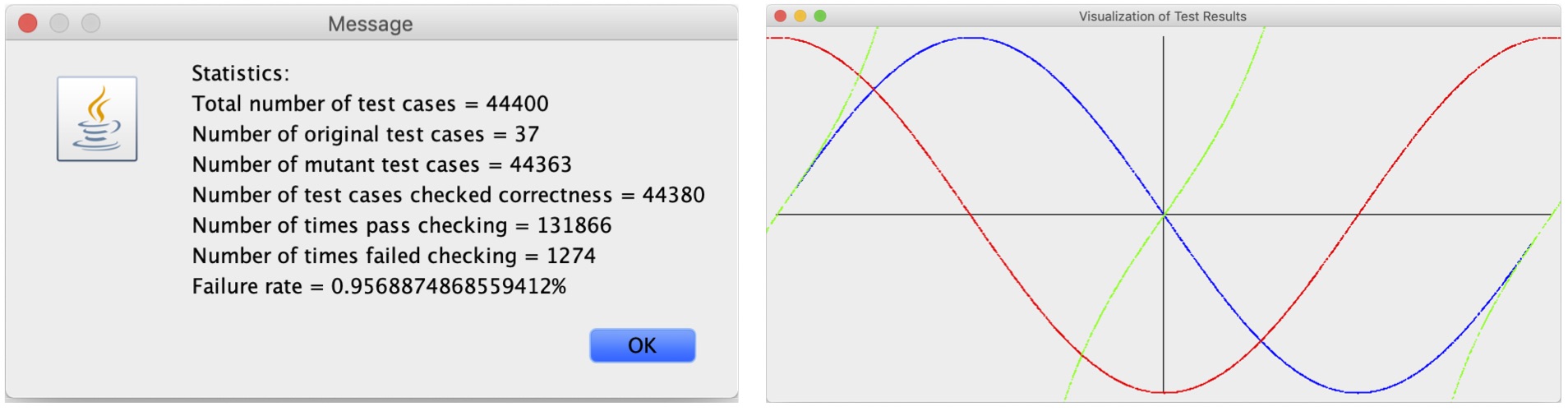}}
\caption{Trigonometric Test Results}
\label{fig:Fig6}
\end{figure}

The second stage starts by generating 20 random test cases in the range of 0 .. $\pi/2$, then applies the datamorphism $x+\pi/2$ and then $x+\pi$. Then the other datamorphisms are applied to populate the test set. The program under test is then executed on the test cases, checked against the metamorphisms, and test results are analysed by invoking the analysers. As shown in Figure \ref{fig:Fig6}, the test on random inputs detected a number of errors. The error rate is $0.957\%$. 

Morphy produces an error message when a test case fails a metamorphism. The following is an example of error message in testing the trigonometric functions.

\begin{footnotesize} 
\begin{verbatim}
-- Rule: tan(pi/2+x) = -1/tan(x) on test case:
{
 id:09f76c14-8852-404e-9865-fac1e73c63a0,
 input:4.71238898038469,
 output:<-1.0|-1.8369701987210297E-16
   |5.443746451065123E15>,
 feature:mutant,
 type:HalfPiPlus,
 origins:[2b954ce1-ad96-488c-a323-0719065eea72],
 correctness:HalfPiPlusSinAssertion=pass;
   HalfPiPlusCosAssertion=pass;
   HalfPiPlusTanAssertion=fail;
}
\end{verbatim}
\end{footnotesize} 

The error messages produced by Morphy show that the errors all happened when the value of tan(x) function is used to checking an identity. Two types of errors have occurred: (a) the value of $tan(x)$ is infinite, e.g. when $x = \pi/2$; (b) the accuracy of an expression is lower than the allowed error, which is $10^{-12}$. 

\SubSection{Face Recognition} 

Face recognition has been employed in the case study on datamorphic testing reported in \cite{Datamorphic2019,DatamorphicTR2019} to demonstrate that datamorphisms can produce meaningful test data for machine learning applications. In this paper, we demonstrate how to achieve test automation and reuses of test code by using Morphy. 

\subsubsection{Reuses of Test Data}

The test data for testing a face recognition application are images of sizes more than 100 KB. In \cite{Datamorphic2019, DatamorphicTR2019}, 200 images of different persons were used and each generated 13 mutants using GAN-Attr \cite{AttGAN} to alter the facial features. Each image is used to test 4 different face recognition applications. A similar experiment was also conducted to test 5 different face recognition applications \cite{SalikThesis2019}. The generation of mutant images from the original images is time consuming. Thus, reusing test data is beneficial. 

To enable the reuse of mutant images, the original images and the mutant images are stored in different folders while they have the same file names. Thus, the image input to a face recognition can be represented by a path to the image file in the test case. The datamorphisms can be written simply as manipulation of strings that takes the path to the original image file and produce that of the mutant image file. The output of a face recognition is a real number expressing the similarity between these two images. 

The following shows the seed maker and one of the datamorphisms for the experiments reported in \cite{DatamorphicTR2019,SalikThesis2019}. It randomly selects a set of original images. Each test case uses one original image as both the target and subject images. 
A datamorphism replaces the subject image with a mutant image. Executing a face recognition application on such a mutant test case examines whether the application recognises the person in mutant images as the same person in the target image. 

\begin{footnotesize} 
\begin{verbatim}
  @MakeSeed
  public void randomSelectOriginalImages(){
    String numStr = JOptionPane.showInputDialog(
      "Entering the number of test cases 
       or \"all\":");
    String[] fileList;
    if (numStr==null) { return;}
    if (numStr.equals("all")) {
      fileList = getFileListAll();
    }else {
      int numTestCases = Integer.valueOf(numStr);
      fileList = getFileList(numTestCases);
    }
    for (String name: fileList) {
      TwoImage input = new TwoImage(
        ImageConfig.imagePath, 
        image+"\\"+name, image+"\\"+name);
      testSuite.addInput(input);
    }
  };
\end{verbatim}
\end{footnotesize} 

Two face recognition experiments are reported in \cite{Datamorphic2019,SalikThesis2019}. We also designed a third experiment which checks if images of different persons can be distinguished by using both the original images and the mutant images. 
The following seed maker generates a number of test cases that each test cases consists of two original images selected at random. 
\begin{footnotesize} 
\begin{verbatim}
  @MakeSeed
  public void randomSelectImagePairs(){
    String numStr = JOptionPane.showInputDialog(
      "Entering the number of test cases:");
    if (numStr==null) { return;};
    int numTestCases = Integer.valueOf(numStr);
    String[] fileList = getFileListAll();
    Random rand = new Random();
    for (int i=0; i<numTestCases; i++) {
      int x = rand.nextInt(fileList.length);
      int y = rand.nextInt(fileList.length);
      String name1 = fileList[x];
      String name2 = fileList[y];
      TwoImage input = new TwoImage(
        ImageConfig.imagePath, 
        image+"\\"+name1, image+"\\"+ name2);
      testSuite.addInput(input);
    }
  };
\end{verbatim}
\end{footnotesize} 

The same set of datamorphisms is applied to these seed test cases using the first order mutant complete strategy. However, the result set of test cases tested a face recognition application on a different property; that is, whether it can distinguish the different persons when using mutant images if using the original images can. Both experiments can be repeated many times by using the seed makers and the datamorphisms without actually regenerating the mutant images many times. In other words, the datamorphisms and the image data are reused for two different types of experiments. 

\subsubsection{Test Executions and Analysis of Results}

Although the applications under test above have different formats of invocation and return messages, the seed makers, datamorphisms as well as other test morphisms can be reused without any change. The only differences are in the test executers. For each face recognition application, a test executer is written to convert a test case object into the invocation request message and then calls the API. It also receives the return message and converts it into a real number representing the confidence in the face recognition. For example, the following invokes Baidu.com face recognition online service. 

\begin{footnotesize} 
\begin{verbatim}
  @TestExecuter
  public Double baiduTest(TwoImage images) {
    ApiTesting apiTesting = new BaiduApi();
    String cmd = images.getCmd();
    try {
      String result = apiTesting.similarity(
        cmd.split(" ")[0], cmd.split(" ")[1]);
      return Double.valueOf(result) / 100;
    } catch (Exception e) {
      e.printStackTrace();
      return 0.0;
    }
  }
\end{verbatim}
\end{footnotesize} 

The face recognition software $SeetaFace$ is an open source project. The code is written in C++. The project is cloned from the GitHub and installed to the local machine where testing is performed. The invocation of the software is through executing a shell command. The following test executer implements the invocation of $SeetaFace$. 

\begin{footnotesize} 
\begin{verbatim}
  @TestExecuter
  public Double seetaFaceTest(TwoImage images) {
    String dir = ImageConfig.seetaFacePath;
    String cmd = dir + File.separator 
      + "Identification.exe " + images.getCmd();
    try {
      Process proc 
        = Runtime.getRuntime().exec(cmd);
      BufferedReader stdout = new BufferedReader(
        new InputStreamReader(
          proc.getInputStream()));
      String result = stdout.readLine();
      stdout.close();
      return Double.valueOf(result);
    } catch (IOException e) {
      e.printStackTrace();
      return 0.0;
    }
  }
\end{verbatim}
\end{footnotesize} 

Two analysers were written to analyse the results of the tests. The following calculates the average of the scores on mutant images for each type of mutant and displays the results on screen. The other saves the data to a file. 

\begin{footnotesize} 
\begin{verbatim}
@Analyser
public void viewStatistics() {
  int numTC = testSuite.testSet.size();
  int numOriginalTC = 0;
  int numMutantTC = 0;
  HashMap<String, List<Double>> resMutant 
    = new HashMap<>();
  for (TestCase<TwoImage, Double> x : 
      testSuite.testSet) {
    if (x.feature == TestDataFeature.original) {
      numOriginalTC++;
    }else {
      numMutantTC++;
      if (!resMutant.keySet().
          contains(x.getType())) {
        resMutant.put(x.getType(), 
           new ArrayList<Double>());
      }
      resMutant.get(x.getType()).add(x.output);
    };
  };
  String message = "Statistics:\n";
  message += "Total number of test cases = " 
      + numTC + "\n";
  message += "Number of original test cases = "
      + numOriginalTC + "\n";
  message += "Number of mutant test cases = " 
      + numMutantTC + "\n";
  for (String type: resMutant.keySet()) {
    Double avg = resMutant.get(type).stream()
      .mapToDouble(Double::doubleValue)
      .average().getAsDouble();
    message += " -- "+ type+" avg = "+avg+"\n";
  }
  JOptionPane.showMessageDialog(null, message);
}
\end{verbatim}
\end{footnotesize} 

\begin{figure}[htbp]
\centering
\scalebox{0.13}{\includegraphics{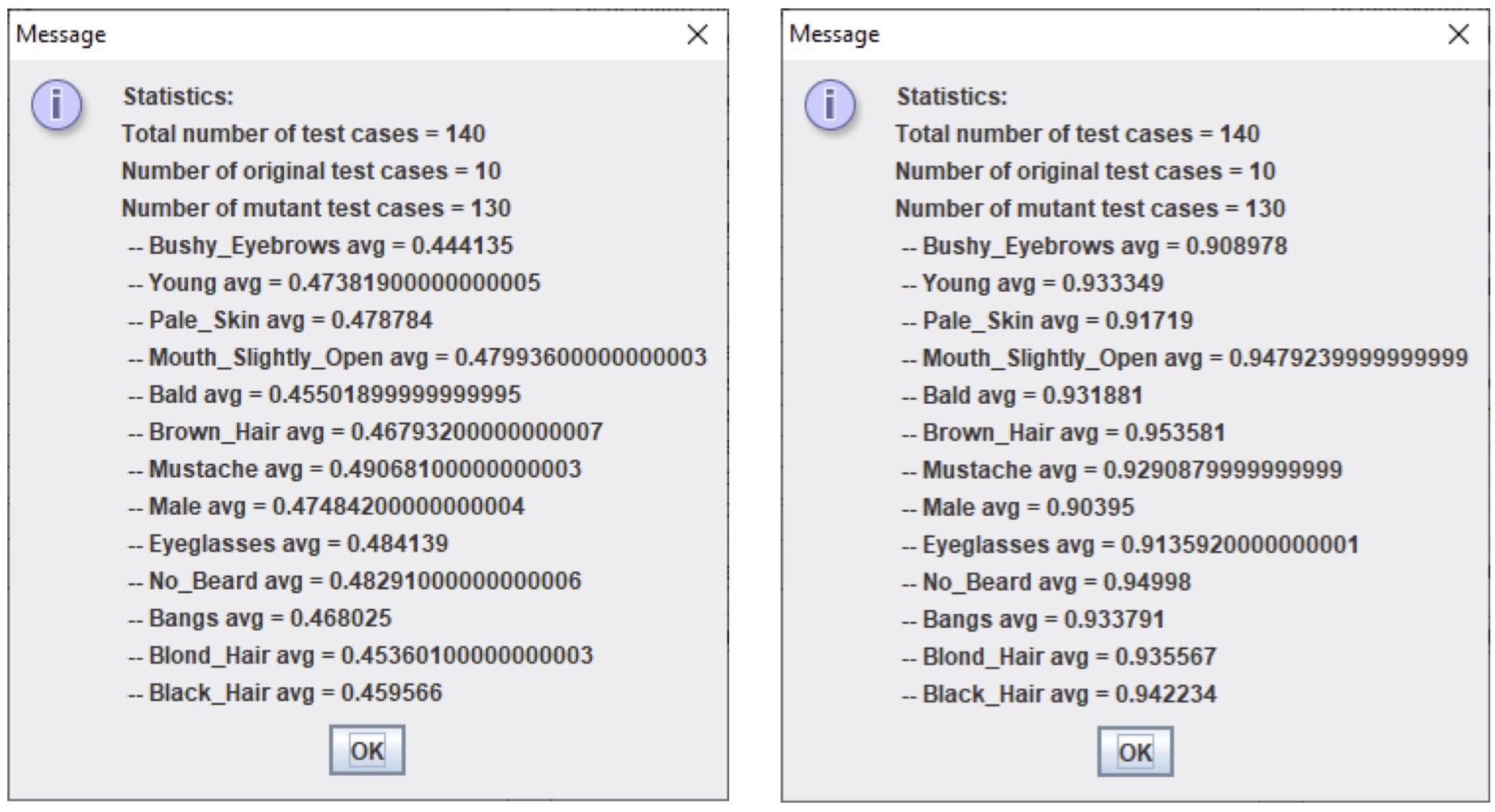}}

{\footnotesize  (a) Images of different persons  ~~~~~~ (b) Images of the same person}
\caption{Face Recognition Test Results}
\label{fig:Fig7}
\end{figure}

Figure \ref{fig:Fig7} gives the screen snapshots of the above analyser on two experiments with one face recognition application. 

\subsubsection{Test Scripts}

As a typical AI application, testing face recognition requires repeated experiments in order to obtain results that are statistically significant. Test scripts were written to further improve test automation for repeated experiments. The following is one of the test scripts written in the case study. 

\begin{footnotesize} 
\begin{verbatim}
//Experiment with images of same person
//Testing Baidu.com face recognition 
saveMessage(ExpDataFile.csv;
  Baidu, same person)
loadTestSpec(file:C:\Users\Hong\git\Morphy3\bin
  \morphy\examples\FaceRecBaiduTest.class)
loadTestSet(C:\Users\Hong\git\Morphy3\ExpData
  \FaceRecognitionExp1)
executeTestCases()
analyse([statisticalAnalysis])

//Testing FacePlus.com face recognition 
saveMessage(ExpDataFile.csv;
  FacePlus, same person)
loadTestSpec(file:C:\Users\Hong\git\Morphy3\bin\
  morphy\examples\FaceRecFacePlusTest.class)
loadTestSet(C:\Users\Hong\git\Morphy3\ExpData
  \FaceRecognitionExp1)
executeTestCases()
analyse([statisticalAnalysis])

//Testing SeetaFace face recognition 
saveMessage(ExpDataFile.csv;
  SeetaFace, same person)
loadTestSpec(file:C:\Users\Hong\git\Morphy3\bin
  \morphy\examples\FaceRecSeetaFaceTest.class)
loadTestSet(C:\Users\Hong\git\Morphy3\ExpData
  \FaceRecognitionExp1)
executeTestCases()
analyse([statisticalAnalysis])

//Testing Tencent.com face recognition 
saveMessage(ExpDataFile.csv;
  Tencent, same person)
loadTestSpec(file:C:\Users\Hong\git\Morphy3\bin
  \morphy\examples\FaceRecTencentTest.class)
loadTestSet(C:\Users\Hong\git\Morphy3\ExpData
  \FaceRecognitionExp1)
executeTestCases()
analyse([statisticalAnalysis])
//Test script end
\end{verbatim}
\end{footnotesize} 

\SubSection{Discussion}

The following observation were made on the case studies. 

First, test morphisms in the case studies are simple and easy to write; see Table \ref{tab:Tab6}, where TC stands for Triangle Classification, Trg for Trigonometric Function, and FR for Face Recognition. LOC is the lines of code. 

\begin{table}[h]
\caption{Summary of Case Studies} \label{tab:Tab6}
\begin{center}
{\footnotesize
\begin{tabular}{|l|c|c|c|}
\hline\hline
                    &TC &Trg &FR\\
\hline\hline
Num of Classes &11 &4 &8\\
\hline
Total LOC &899 &830 & 450\\
\hline
Num of Seed Makers & 4 & 3 &3\\
\hline
Average LOC of Seed Makers &26.25 &61.67 &21.33\\
\hline
Num of Datamorphisms &20 &10 &13\\
\hline
Average LOC of Datamorphisms &9 &6 &8\\
\hline
Num of Metamorphisms &25 & 30 & --\\
\hline
Average LOC of Metamorphisms &8.72 &7.00 &--\\
\hline
Num of Analysers &2 & 2 &2\\
\hline
Average LOC of Analysers &62 &33 &41\\
\hline\hline
\end{tabular}
}
\end{center}
\end{table}%

Second, test specifications can be reusable if it is decomposed into a number of classes where common test morphisms are placed together while test specific morphisms are placed in separate classes. 

Third, achieving test automation using facilities at three different levels of activity, strategy and process is flexible and practical. Different testing tools and techniques can be easily integrated into the Morphy framework and used together. 

\Section{Conclusion} \label{sec:Sec5}
\SubSection{Main Contributions}
This paper redefines the datamorphic testing method by classifying test artefacts into test entities and test morphisms. The notion of test morphisms is defined as mappings between software test entities. Datamorphisms and metamorphisms are examples of such test morphisms. The former are mappings from test cases to test cases, while the later are mapping from test cases to Boolean values as assertions on the correctness of the test. Seed makers are also test morphisms, which map into test sets. We identified a set of other test morphisms, which include test case metrics and filters, test set metrics and filters, test executers and analysers.

The paper also presents a test automation tool called $Morphy$. We demonstrated that testing activities can be automated by writing test codes for various test morphisms and invoking such test codes through a test tool like Morphy. Advanced combinations of test morphisms can be realised by implementing of test strategies to achieve a higher level of test automation. Three types of test strategies have been implemented in Morphy: (a) Datamorphism combination strategies, which generate test sets of various coverage of datamorphisms combinations; (b) Exploration strategies, which explore the test space in order to find the borders between subdomains for testing classification and clustering type of AI applications; (c) Test set optimisation strategies, which employ genetic algorithms optimise test sets. This paper focuses on the first of these, defining both strategies and implementation algorithms, with the other types of test strategies being reported separately. 

Morphy also provides a record-replay type of test automation facility to further improve test automation especially for regression testing. Test scripts can be recorded from the interactive invocations of test morphisms for basic test activities and invocations of test strategies as well as test management activities such as load test specifications and load and save test sets, etc. Although the case studies reported in this paper used test scripts to improve test automation, a more detailed study of the test script facility will be reported in a separate paper. 

\SubSection{Related Work}

Morphy is a test automation framework applicable to all kinds of software systems including AI applications as demonstrated in the case studies. In comparison with existing test automation frameworks like XUnit \cite{UnitTestFranmeworkBook, xUnitPattersBook2007} and GUI test script based test tools like Selenium \cite{SeleniumWebHome} and WebDriver \cite{WebDriverStandard2018}, Morphy provides more advanced test automation facilities such as test strategies and test scripting. 

In XUnit framework, testing is defined by a set of methods in a class or a set of test scripts for executing the program under test together with methods for setting up the environment before test executions and tearing down the environment after test. Such a test specification is imperative. Our test specification is declaratively imperative: various types of test morphisms are declared in a test specification while each test morphism is coded in an imperative programming language. Our case studies show that such test specifications are highly reusable and composable even for testing different applications. This is what existing test automation frameworks have not achieved. 


In a GUI based test automation framework, test automation is realised by test scripts or test code that interact with GUI elements. The most representative and most well-known example of such testing tools is Selenium \cite{SeleniumWebHome}. Selenium has two test environments: (a) the Selenium IDE in which manual GUI based testing can be recorded into a test script and replayed; (b) the Selenium Web Drivers, which provides an API with methods for web-based GUI elements. Test code can be written in various programming languages that are equivalent to test scripts. Morphy also employs test scripts to achieve a high level of test automation, but it is equipped with more advanced test automation facilities such as test strategies. 

An advantage of Morphy is that the architecture enables various testing techniques and tools to be integrated by wrapping exiting testing tools as methods in test specification class that invoke the tools. For example, test case generation techniques and tools \cite{TestGenerationOrchestratedSurvey2013} like fuzz testing \cite{FuzzTestBook2007}, data mutation testing \cite{DataMutationJournal2009}, random testing \cite{RandomTestingSurvey2012}, adaptive random testing \cite{AdaptiveRandomTestTY2004, AdaptiveTestZhu2008}, combinatorial testing \cite{CombinattorialTestingSurveyNie2011} and model based test case generators are all test morphisms, which can be wrapped as seed makers or datamorphisms. Metamorphic relations in metamorphic testing \cite{MetamorphicTestingSurveyChen2018} and formal specification-based test oracles \cite{BGM91} are metamorphisms. Algebraic specification has been used for both generating test cases and checking test correctness \cite{BlackAndWhite98, TACCLE2001, AlgebraicTestOracleZhu2003}. The techniques that implements automated testing based on algebraic specifications \cite{BGM91, KongZZ07, MonicTest2016} can easily be split into two parts, the test morphisms to generate test cases and the test morphisms to check test correctness. Test coverage measurement tools like \cite{CoverageToolListWebpage2017} are test set metrics. Regression testing techniques and methods \cite{RegressionTestingSurveyYoo2012} that select or prioritise test cases in an existing test set can be implemented as test set filters. Search-based testing \cite{SearchBasedSESurveyHarman2012, SearchBaseTestingSurvey2015} can be regarded as test strategies. Therefore, they can all be easily integrated into Morphy. 

\SubSection{Future Work}

It is worth noting that datamorphic testing focuses on test morphisms related to test data and test sets, as its name implies. There are other types of test morphisms. For example, mutation operators in mutation testing \cite{MutationTestingSurveyHarman2011} and fault injection tools for other fault-based testing methods are test morphisms that are mappings from programs to programs or sets of programs. Specification mutation operators are test morphisms that mapping from formal specifications to specifications or sets of specifications. It is an interesting further research question how to integrate such test morphisms into the datamorphic testing tools like Morphy, although, theoretically speaking, there should be no significant difficulty to do so. 

It is also possible to integrate XUnit test automation frameworks like JUnit and GUI based test automation tools like WebDriver with Morphy. This is also an interesting topic for future work. 

We have already conducted some experiments with the exploratory strategies and genetic algorithms for test set optimisation strategies. The results will be reported in separate papers. 

%

%

\newpage
\onecolumn
\begin{figure}[htbp]
\scalebox{1}{\includegraphics{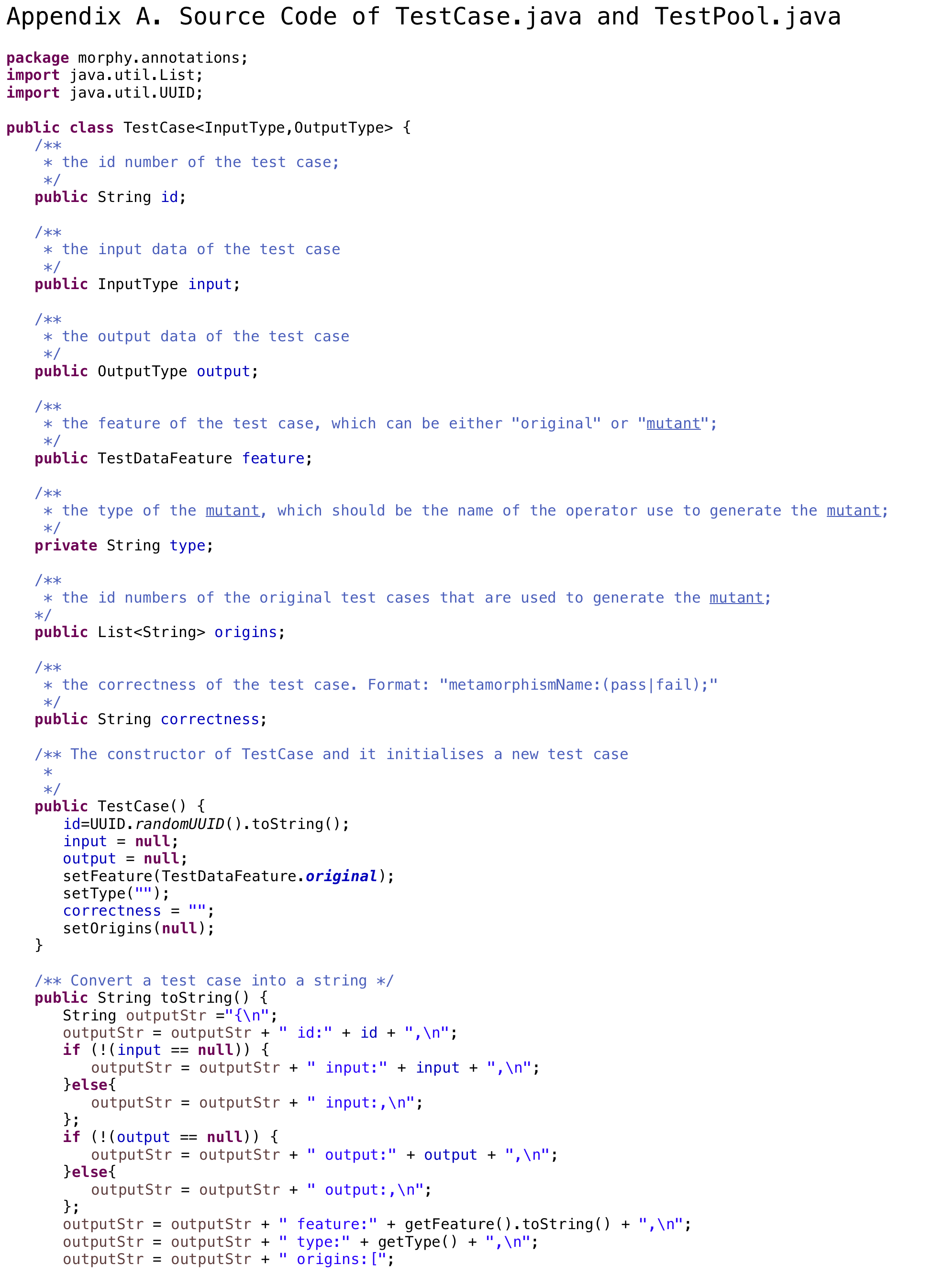}}
\end{figure}

\begin{figure}[htbp]
\scalebox{1}{\includegraphics{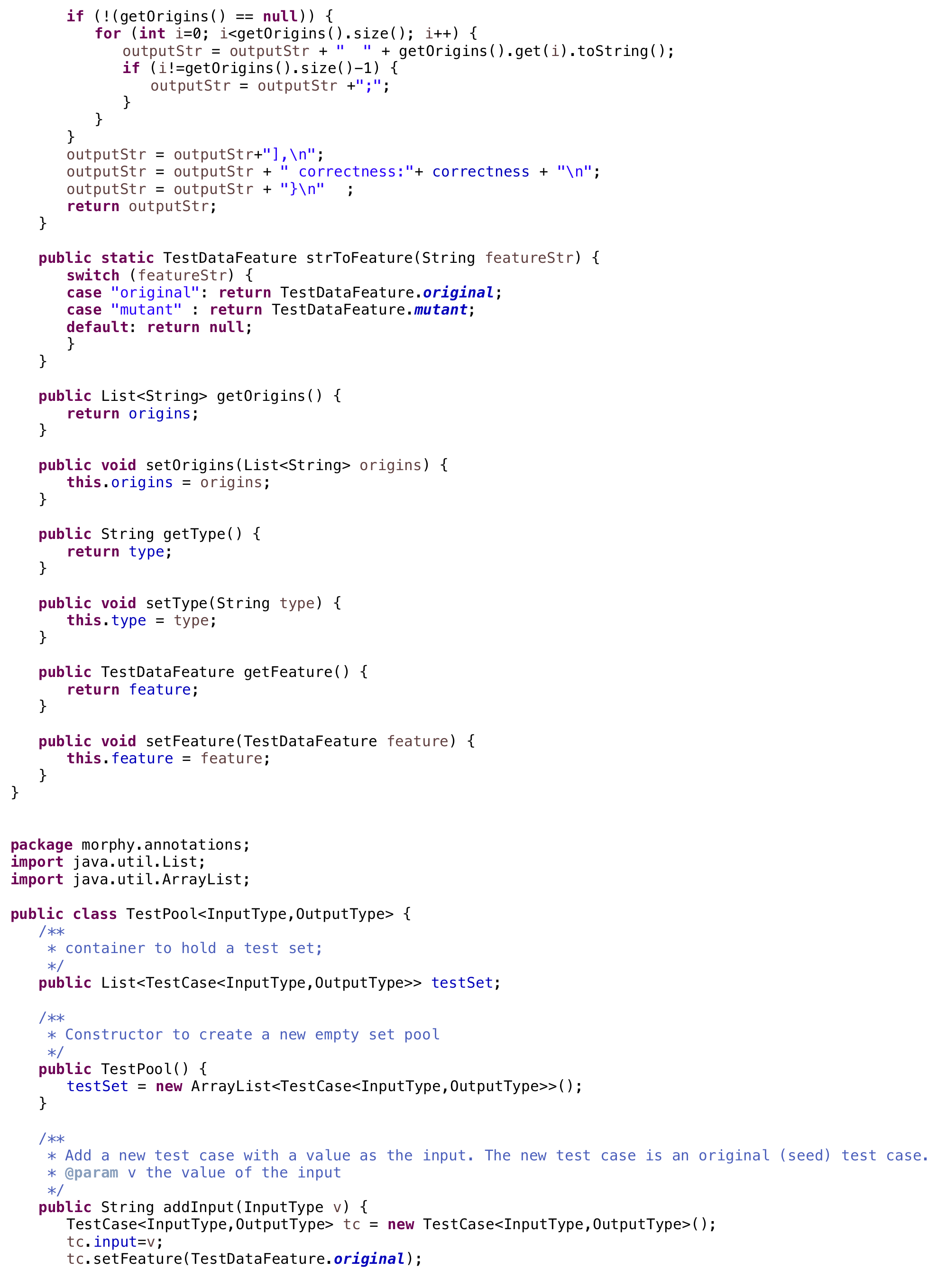}}
\end{figure}

\begin{figure}[htbp]
\scalebox{1}{\includegraphics{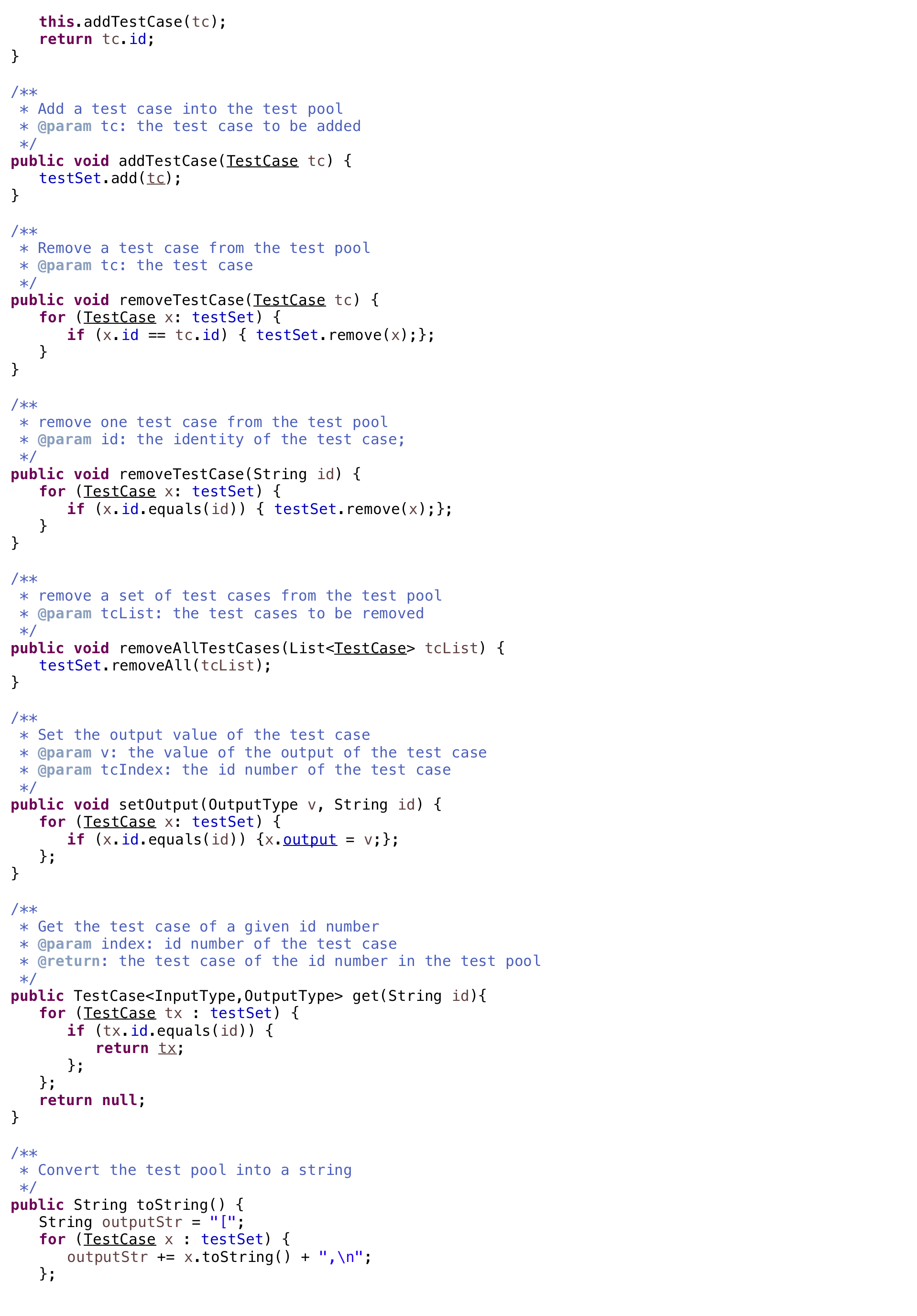}}
\end{figure}

\begin{figure}[htbp]
\scalebox{1}{\includegraphics{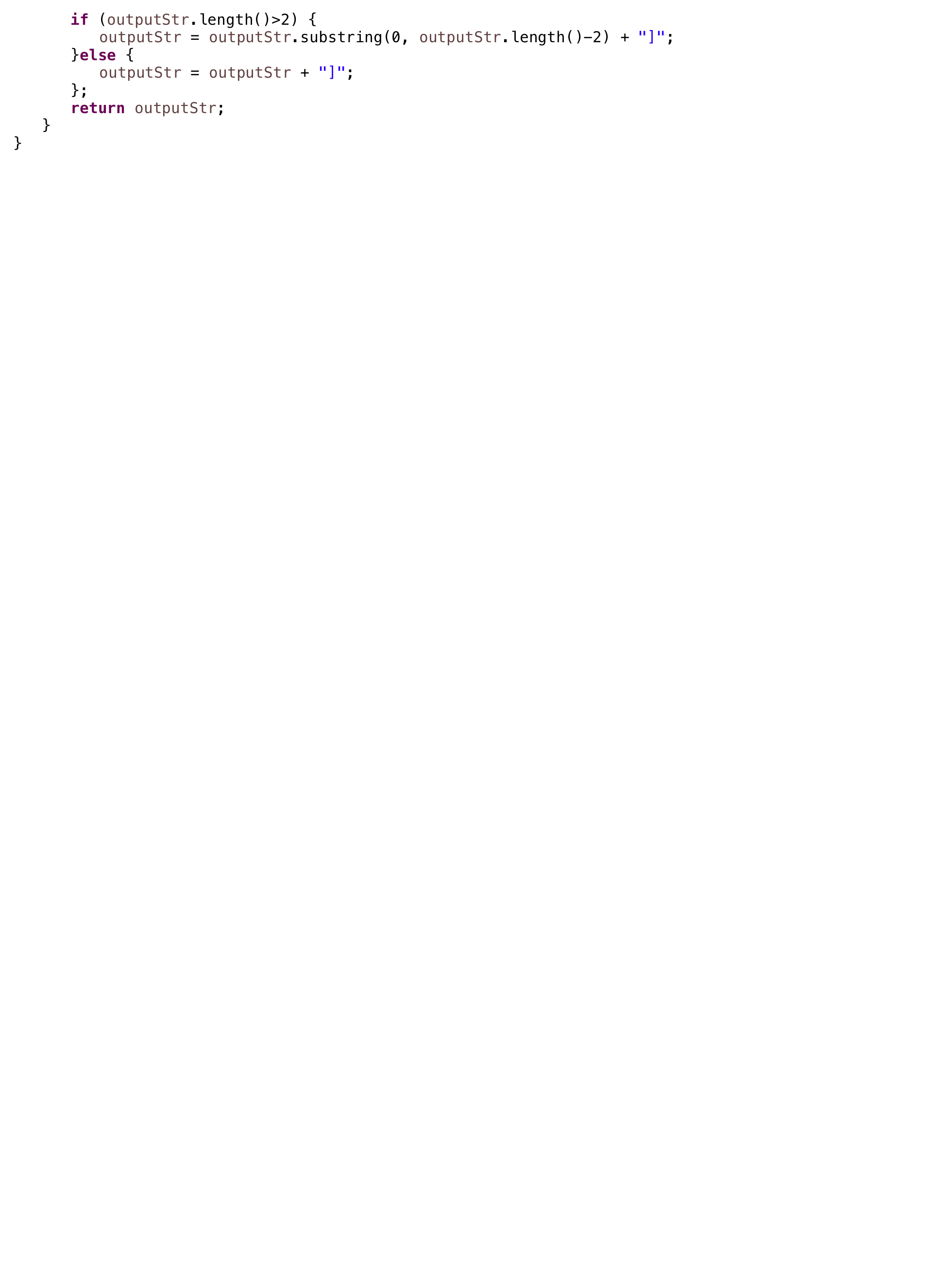}}
\end{figure}


\begin{figure}[htbp]
\scalebox{1}{\includegraphics{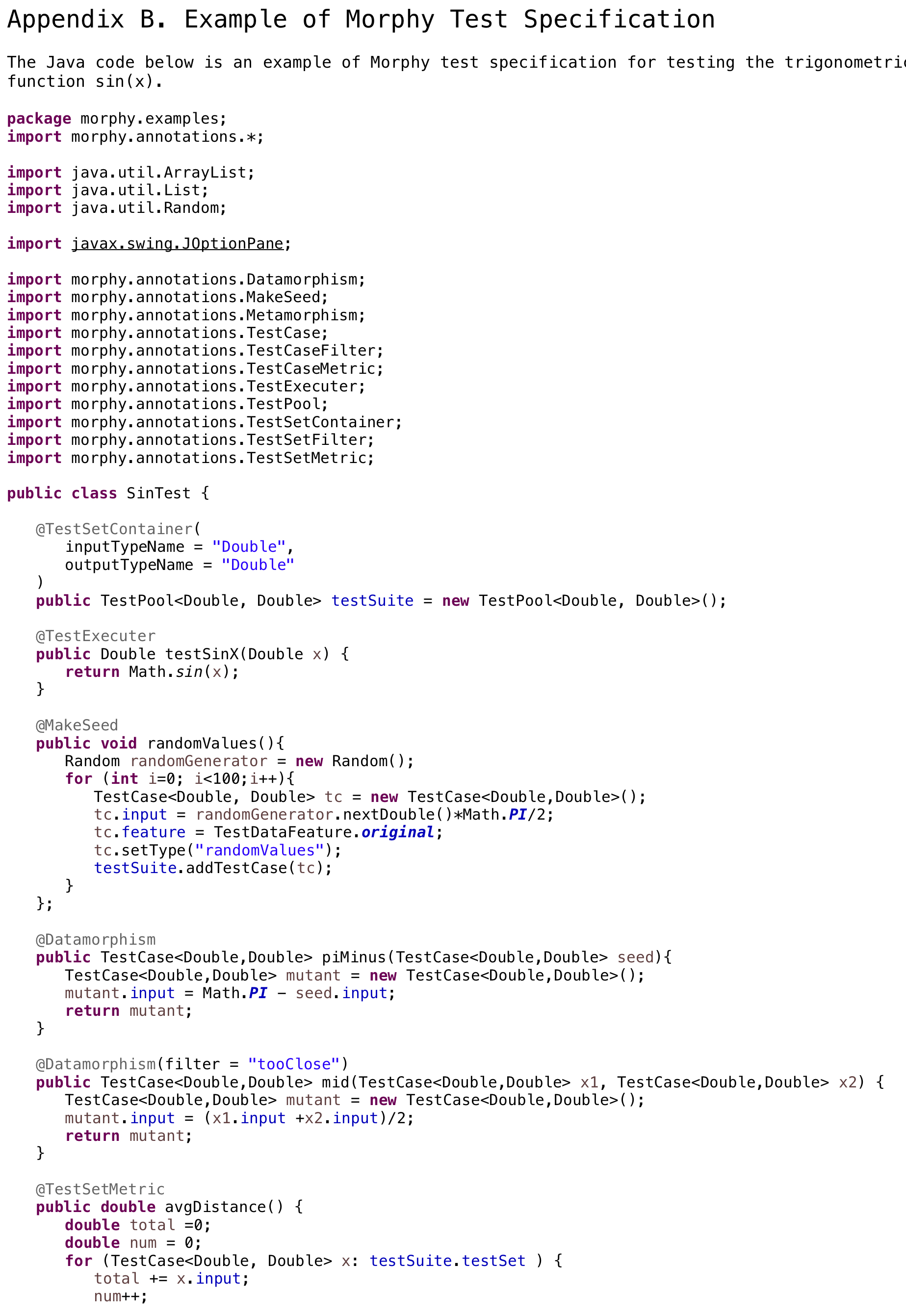}}
\end{figure}

\begin{figure}[htbp]
\scalebox{1}{\includegraphics{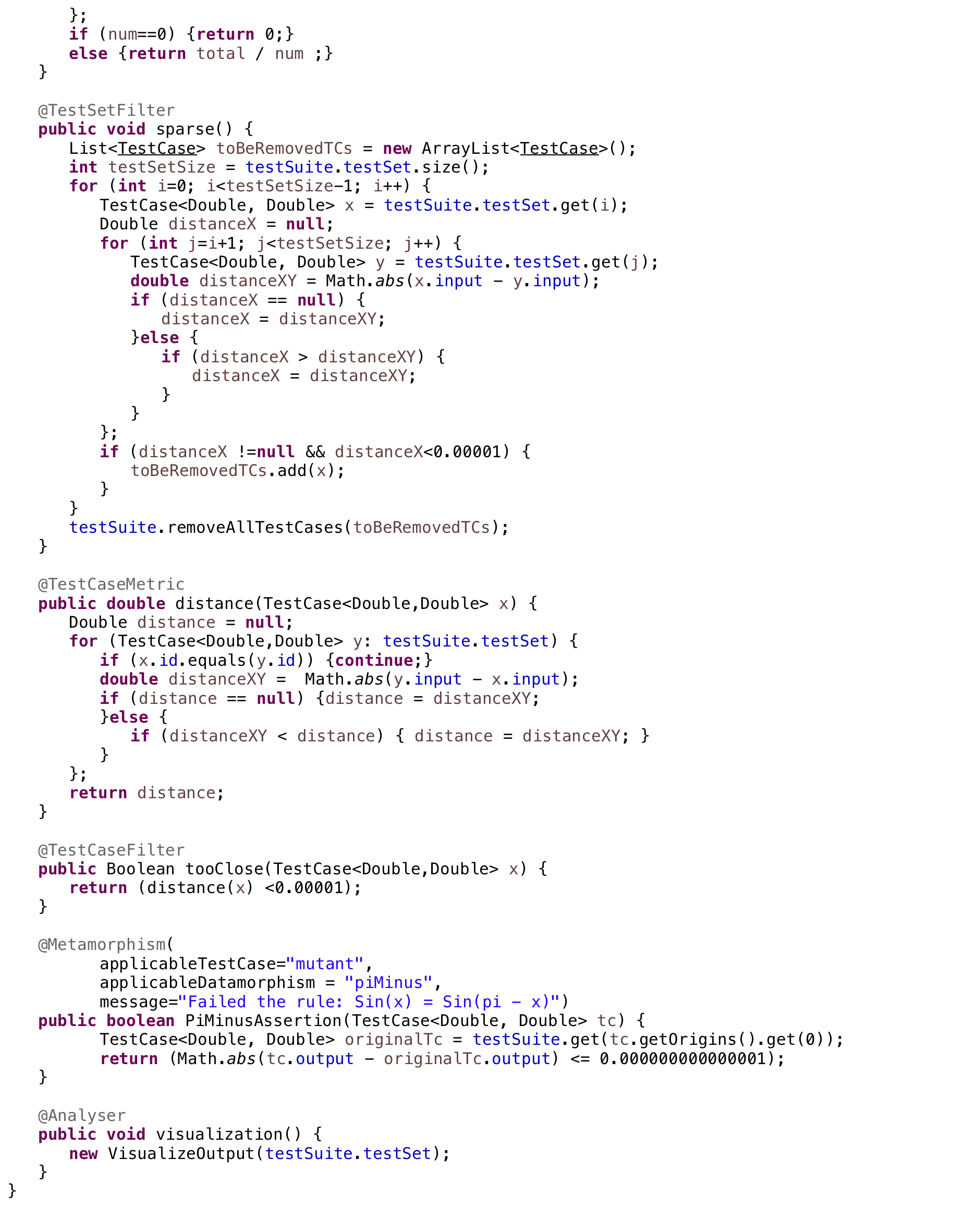}}
\end{figure}

\begin{figure}[htbp]
\scalebox{1}{\includegraphics{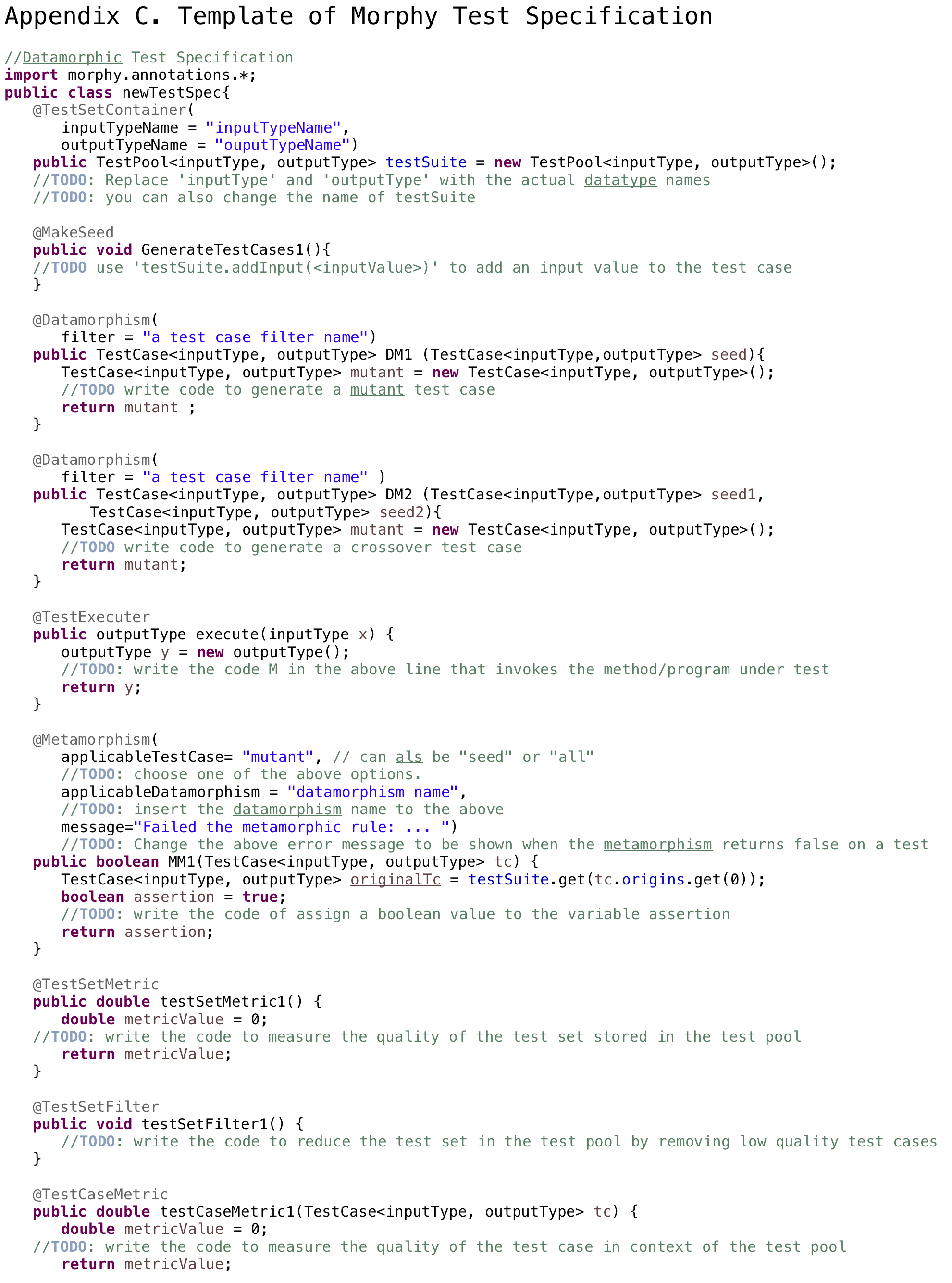}}
\end{figure}

\begin{figure}[htbp]
\scalebox{1}{\includegraphics{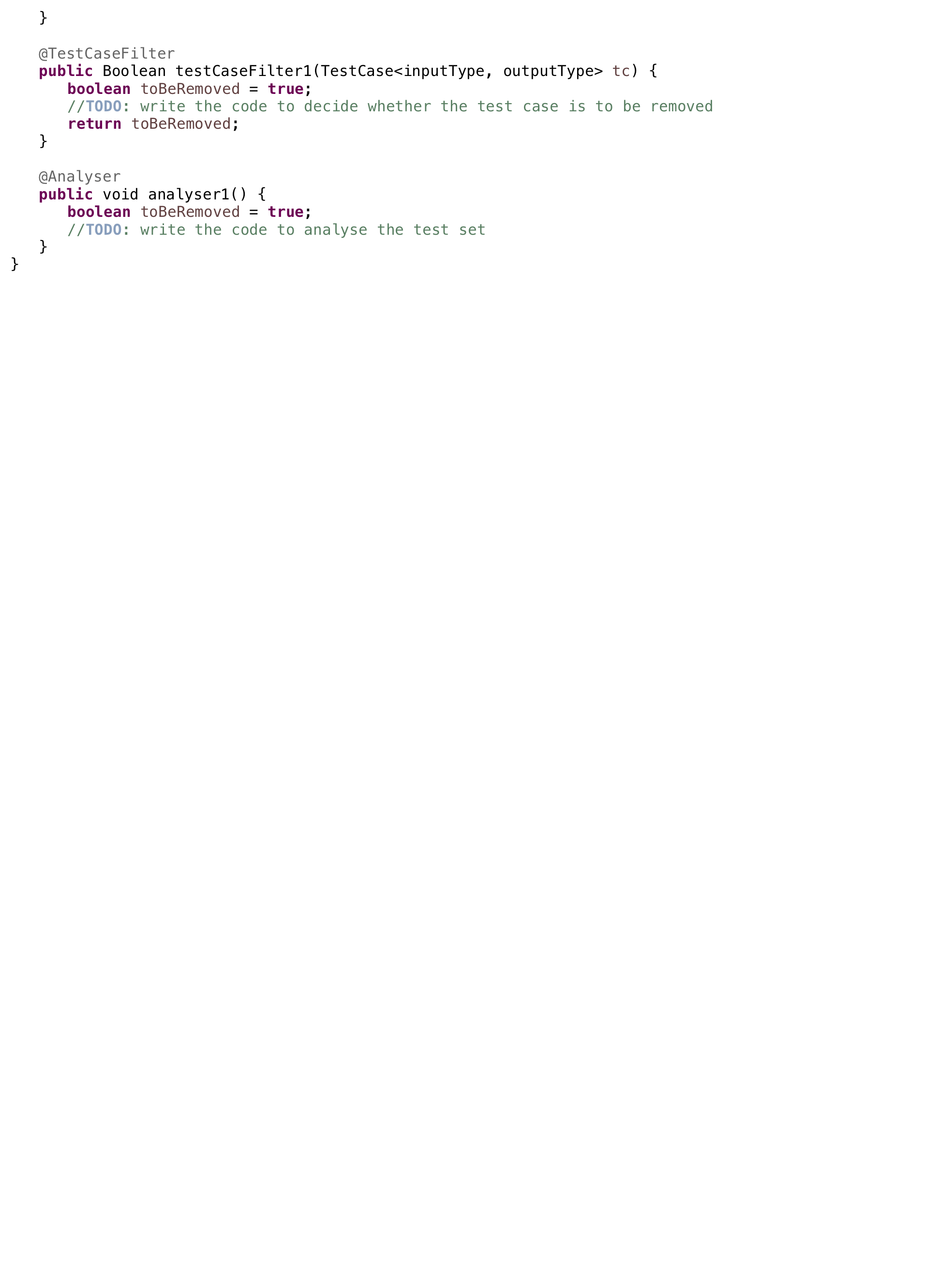}}
\end{figure}


\end{document}